\magnification1200

\rightline{KCL-MTH-11-20}

\vskip .5cm
\centerline {\bf Higher derivative type II string effective
actions, automorphic forms and $E_{11}$ }

\vskip 1cm
\centerline{Finn Gubay and Peter West}
\centerline{Department of Mathematics}
\centerline{King's College, London WC2R 2LS, UK}

\vskip .5 cm
\noindent
By dimensionally reducing the ten-dimensional higher derivative
type IIA string theory effective action we place constraints on the
automorphic forms that appear in the effective action in lower
dimensions.
We propose Êa number of Êproperties of such automorphic forms and
consider the prospects that $E_{11}$ can play a role in the formulation
of the higher derivative string theory effective action.
\vfill
\eject

\noindent
{\bf {1. Introduction}}
\bigskip
The low-energy effective actions of the IIA and IIB string theories are
the IIA [1-3] and IIB [4-6] supergravity theories. ÊFurthermore
eleven-dimensional supergravity [7] is the low-energy
effective action of one of the limits of M-theory. The type IIA and
type IIB supergravity theories contain all
perturbative and non-perturbative string effects and as a consequence
their study has lead to many aspects of what we now know about
string theory. Upon dimensional reduction of the IIA and IIB theories on
an $n$ torus, or equivalently the eleven dimensional theory on an $n+1$
torus, to
$d = 10 - n$ dimensions all these theories become equivalent and
possess a
hidden
$E_{n+1}$ duality symmetry [8-11]. The IIB supergravity theory also
possesses an $SL(2,R)$ symmetry [4]. The four-dimensional heterotic
supergravity theory possesses an analogous $SL(2,R)$ symmetry and taking
into account the fact that the brane charges are quantised [12,13] and
rotated by this symmetry it was proposed [14,15] that the four
dimensional heterotic string theory Êwas invariant under an $SL(2,Z)$
symmetry which included a transformation that mixed perturbative to
non-perturbative effects. This realisation was generalised to the
$E_{n+1}$ symmetry of type II theories in [16].
\par
The higher derivative corrections to string theory have
been most studied in the context of IIB string theory where it was
found that demanding that the theory is Êinvariant under the $SL(2,Z)$
symmetry Êleads to the appearance of
automorphic forms that place very strong constraints on the theory
[17-24]. ÊFor type IIB string theory compactified to eight or nine
dimensions, invariance under the corresponding U-duality groups
similarly
lead to the appearance of automorphic forms [25-29]. ÊThe role of
automorphic forms in the low energy effective action of type II string
theory was Êalso Êdiscussed sometime ago in seven and fewer dimensions
[30,31]. More recently the higher derivative corrections of type II
string theories in less than ten dimensions, including dimensions less
than seven, have been systematically studied [32-39] and specific
automorphic forms have been proposed for certain higher derivative terms
constructed from particular representations of
$E_{n+1}$. Furthermore the regulation Êof the divergences was
carried out and precise ÊÊÊpredictions for the perturbative series
worked
out in detail [34-38]. ÊIn particular these papers have generalised the
previous results in non-renormalisation theorems [25,28,40].
These studies have, however, Êbeen limited to terms with relatively
lower numbers of space-time derivatives and very little is known about
such terms Êin general. An exception was that of Êreference [39] in
which the dimensional reduction of arbitrary higher derivative terms in
the IIB theory on an
$n$ torus were compared with the result expected in
$10-n$ dimensions if an $E_{n+1}(Z)$ symmetry was present. In this way
one was able to place some restrictions on the representations used to
construct the automorphic forms for an arbitrary higher derivative
correction. A similar analysis was also carried out but starting from
eleven dimensions.

In this paper we will follow a similar approach to that of reference
[39],
but from the view point of the IIA theory. In particular we will
consider
the dimensional reduction Êof the higher derivative string corrections
of the IIA theory on an $n$-dimensional Êtorus Êto $d=10-n$ dimensions.
We will compare these with the higher derivative corrections that arise
in the $d$ dimensional theory assuming that the theory is invariant
under an $E_{n+1}(Z)$ symmetry and so possess a corresponding
automorphic
form built from a representation of $E_{n+1}(Z)$.
This comparison allows us to ÊÊplace
constraints on the representation used to construct the automorphic form
that appears for any higher derivative correction.
Indeed we find that the highest weight
$\vec{\Lambda}_{n}$ Êappears Êin the Êautomorphic form,
where $\vec{\Lambda}_{n}$ is the highest weight of the fundamental
representations of $E_{n+1}$ associated with node Ê$n$. The Dynkin
diagram
of $E_{n+1}$ with the labelling of the nodes is given in figure
five. This strongly suggests that each higher ÊÊÊderivative
correction contains an automorphic form constructed ÊÊfrom this
fundamental representation.
\par
In order to carry out the comparison we need to identify the fields that
arise in the dimensional reduction from ten dimensions with the fields
that occur in the formulation of the $d$-dimensional theory in which the
$E_{n+1}(Z)$ symmetry is manifest, and in particular the scalar fields
from which the automorphic form is constructed. This identification
can be
carried out in the context of the supergravity theories. The most
obvious
technique is to explicitly carry out the dimensional reduction of the
supergravity theory and reformulate the theory with the manifest
$E_{n+1}(Z)$ symmetry, but this is rather lengthy and complicated
involving dualisations and other subtleties.
In this paper we will use the
$E_{11}$ formulation of the IIA theory [43,55]. In this formulation the
fields of the
theory are in one to one correspondence with the generators of the
Borel subalgebra of $E_{11}$. As the Ê$E_{11}$ algebra contains in an
obvious way the $E_{n+1}$ algebra, the correspondence between the
scalar fields that appear in the non-linear realisation of $E_{n+1}$ and
the $E_{11}$ generators is easily found. However, the correspondence
between ÊÊthe $E_{11}$ generators and the fields usually used to
formulate the IIA supergravity theory is known from the formulation of
this theory as a non-linear realisation at lowest levels in $E_{11}$
[41,43].
Thus
one finds the desired relation between the fields of the IIA theory and
the scalars fields ÊÊÊassociated ÊÊÊwith ÊÊ$E_{n+1}$ in a simple way.
We note that although there is strong evidence for the conjecture
that $E_{11}$ is an underlying symmetry of the theory of strings and
branes our use of $E_{11}$ in this paper does not rely upon the $E_
{11}$ conjecture holding.
\par
After discussing the consequences of the results of this paper we make
a number of proposals for the properties of the automorphic forms that
occur in string theory for any number of space-time derivatives. We
also consider the possibility that the higher derivative effective
action admits an $E_{11}$ formulation.

\bigskip
\noindent
{\bf {2. The Dimensional Reduction}}
\bigskip
The bosonic field content of type IIA supergravity in ten dimensions
consists of a scalar (the type IIA dilaton $\phi$), a NS-NS three form
field strength $\tilde{F}_{\mu_{1} \mu_{2} \mu_{3}}$ constructed from
the
NS-NS two form gauge field $\tilde A_{\mu_{1} \mu_{2}}$, in addition to
two R-R
form field strengths $\tilde{F}_{\mu_{1} \mu_{2}}$, $\tilde{F}_{\mu_{1}
\mu_{2} \mu_{3} \mu_{4}}$ constructed from the R-R
gauge fields $\tilde A_{\mu_{1}}$ and $ \tilde A_{\mu_{1} \mu_{2}
\mu_{3}}$. In Einstein
frame, the bosonic part of the type IIA supergravity action is given by
[1,2,3],
$$
S_{IIA}= {1 \over 2\kappa_{10}^{2}} Ê\int d^{10}x
det{\left(\tilde{e}\right)}\left( \tilde{R} - {1 \over 2 \cdot 4!} e^
{ {1
\over 2} \phi} \tilde{F}_{\mu_{1} \mu_{2} \mu_{3} \mu_{4}}
\tilde{F}^{\mu_{1} \mu_{2} \mu_{3} \mu_{4}} - {1 \over 2 \cdot 3!}
e^{-\phi} \tilde{F}_{\mu_{1} \mu_{2} \mu_{3} } Ê\tilde{F}^{\mu_{1}
\mu_{2}
\mu_{3} } \right.
$$
$$
\left. - {1 \over 2 \cdot 2!} e^{{3 \over 2} \phi} \tilde{F}_{\mu_{1}
\mu_{2}} \tilde{F}^{\mu_{1} \mu_{2}} - {1 \over 2} \partial_{\mu}
\phi \partial^{\mu} \phi \right),
\eqno(2.1)
$$
where Ê$\kappa_{10}$ is a constant related to the Newton constant in
ten dimensions and
$$
\tilde{F}_{\mu_{1} \mu_{2} } = Ê2 \partial_{ [ \mu_{1}} \tilde{A}_
{\mu_{2} ]
},
$$
$$
\tilde{F}_{\mu_{1} \mu_{2} Ê\mu_{3} } = Ê3 \partial_{ [ \mu_{1}}
\tilde{A}_{\mu_{2} \mu_{3} ] },
$$
$$
\tilde{F}_{\mu_{1} \mu_{2} Ê\mu_{3} \mu_{4} }= Ê4\left( \partial_{ [
\mu_{1}} \tilde{A}_{\mu_{2} \mu_{3} \mu_{4} ] } + \tilde{A}_{[ \mu_{1} }
\tilde{F}_{\mu_{2} \mu_{3} \mu_{4} ] } \right).
\eqno(2.2)$$
We have suppressed Êthe Chern-Simons term since it will not
play a part in our analysis. ÊThe type IIA supergravity action possesses
a $GL(1,R)$ symmetry, that manifests itself through a shift of the type
IIA dilaton and a scaling of the other fields. One can introduce
the following combinations of the field strengths and dilaton that are
inert under
$GL(1,R)$ transformations
$$
\tilde{\cal{F}}_{\mu_{1} \mu_{2} \mu_{3} \mu_{4}}=e^{{1 \over 4}\phi}
\tilde{F}_{\mu_{1} \mu_{2} Ê\mu_{3} \mu_{4} },
$$
$$
\tilde{\cal{F}}_{\mu_{1} \mu_{2} \mu_{3}}=e^{-{1 \over 2}\phi} \tilde
{F}_{\mu_{1} \mu_{2} Ê\mu_{3} },
$$
$$
\tilde{\cal{F}}_{\mu_{1} \mu_{2}}=e^{{3 \over 4}\phi} \tilde{F}_{\mu_
{1} \mu_{2} }.
\eqno(2.3) $$
In fact these are just the non-linear representations of
$GL(1,R)$ constructed from the linear representations in the usual way
(see appendix A). They are inert, as the local subalgebra is the
identity
group. Rewriting the action with these objects effectively absorbs the
dilaton factors multiplying the field strengths in (2.1), the action
then becomes
$$
S_{IIA}= {1 \over 2\kappa_{10}^{2}} Ê\int d^{10}x
det{\left(\tilde{e}\right)}\left( \tilde{R} - {1 \over 2 \cdot 4!}
\tilde{\cal{F}}_{\mu_{1} \mu_{2} \mu_{3} \mu_{4}}
\tilde{\cal{F}}^{\mu_{1} \mu_{2} \mu_{3} \mu_{4}} - {1 \over 2 \cdot 3!}
\tilde{\cal{F}}_{\mu_{1} \mu_{2} \mu_{3} } Ê\tilde{\cal{F}}^{\mu_{1}
\mu_{2} \mu_{3} } \right.
$$
$$
\left. - {1 \over 2 \cdot 2!} Ê\tilde{\cal{F}}_{\mu_{1} \mu_{2}}
\tilde{\cal{F}}^{\mu_{1} \mu_{2}} - {1 \over 2} \partial_{\mu} \phi
\partial^{\mu} \phi \right).
\eqno(2.4) $$
\par
In this paper we are interested in the dimensional reduction of a
generic ten dimensional type IIA higher derivative term which may be
written as
$$
\int d^{10} x det (\tilde{e}) {\partial}^{\tilde{l}_{0}} \tilde{R}^{{
\tilde{l}_{R}}
\over 2} (\tilde{P}_{\mu_{1}} )^{\tilde{l}_{1}}
(\tilde{\cal F}_{\mu_1\mu_2})^{{\tilde{l}_2}}
\tilde{\cal F}_{\mu_1
\mu_2 \mu_3})^{\tilde{l}_{3}} Ê(\tilde Ê{\cal F}_{\mu_1
\mu_2 \mu_3 \mu_4})^{\tilde{l}_{4}} \Phi_{\tilde{s}},
\eqno(2.5) $$
where $\Phi_{\tilde{s}}$ is a function of $\phi$ that is of the form
$\Phi_{\hat{s}}=e^{-\tilde{s} \phi}$.
Dimensional
reduction on an
$n$ torus to a theory in $d=10-n$ dimensions is achieved using the
metric compactification ansatz
$$
d\tilde {s}^{2}_{10}=e^{2\alpha \rho}g_{\mu
\nu}dx^{\mu}{dx^{\nu}}+e^{2 \beta
\rho}G_{ij}\left(dx^{i}+A^{i}_{\mu}dx^{\mu} \right)\left(dx^{j}+A^{j}_{
\nu} dx^{\nu} \right),
\eqno(2.6)$$
where the background and internal metrics are denoted $g_{\mu \nu}$ and
$G_{ij}$ respectively, with the internal metric satisfying $det(G)=1$
and
$$
\alpha=\sqrt{{n \over 2\left( d-2 \right) \left( D-2 \right)} }, \ \
\beta=-{\left(d-2\right) \alpha \over n}.
\eqno(2.7)$$
For us in this paper $D=10$. The internal vielbein is
given by $
e_{i}{}^{\underline k} e_{j}{}^{\underline l}
\delta_{\underline k\underline l} = G_{ij}$ and satisfies $\det (e)
=1$.
Tangent internal indices possess an underline as shown. The gauge fields
are dimensionally reduced in the obvious way using world indices i.e.
$\tilde A_{\hat \mu}, \ \hat \mu=0,\ldots , 9$ is set equal Êto $A_{i}
$ if
$
\hat
\mu=i, i=d+1, \ldots ,9$ and
$A_{\mu}$ if
$\hat \mu=\mu, \ \mu=0,\ldots ,d$.
\par
We will be interested in the dependence of the above ten dimensional
higher derivative correction in string frame. The transition from
Einstein frame to string
frame is given by $\tilde{e} = e^{-{\phi \over 4}}\tilde{e}_{s}$.
The term in (2.5) then leads to the factor
$$
e^{{\phi \over 4} \left( \tilde{l}_{0} + \tilde{l}_{R} + \tilde{l}_{1}
+ 5 \tilde{l}_{2}+ \tilde{l}_{3}+ 5 \tilde{l}_{4} - 10 - 4 \tilde
{s} ÊÊÊ\right) Ê}.
\eqno(2.8)
$$
At order $g$
in perturbation theory we have the contribution $e^{\phi(2g-2)}$ and
so for a
perturbative contribution we find
$$
\tilde{s}= {1 \over 4} \left( Ê\tilde{l}_{0} + \tilde{l}_{R} + \tilde
{l}_{1}+ 5 \tilde{l}_{2}+ \tilde{l}_{3}+ 5 \tilde{l}_{4} - 2 - 8g
\right).
\eqno(2.9)
$$
\par
The Êdimensionally reduced theory will Êcontain field strengths
of the form $F_{\mu_{1}...\mu_{p} i_{1}...i_{k}}$ where the internal
indices $i_1, \ldots, i_k $ are world volume indices. The theory in $d$
dimensions possesses the $GL(1,R)$ symmetry of the IIA theory, but in
addition has an $SL(n,Z)$ symmetry corresponding to diffeomorphisms that
are preserved by the torus. We can convert the internal world indices to
tangent frame indices using the inverse internal vielbein. However, as
explained in reference [39] page 5, the internal vielbein is just the
group element of the non-linear realisation of $SL(n)$ with local
subgroup
$SO(n)$ in the vector representation. Carrying this out on Êthe
$GL(1,R)$ inert objects of equation (2.3) we find
${\cal F}_{Sl(n)\otimes GL(1)\ j_{1}...j_{k}}$, Êwith any space-time
indices suppressed, which converts to tangent space as follows [32]
$$
{\cal F}_{Sl(n)\otimes GL(1)\ \underline{i}_{1}...\underline{i}_{k}} Ê=
(e^{-1})_{\underline{i}_{1}}^{\
\ j_{1}}... (e^{-1})_{\underline{i}_{k}}^{\ \ j_{k}} ÊF_{j_{1}...j_{k}},
\eqno(2.10)$$
where $e_i{}^{\underline k}$ is the vielbein on the torus and
$F_{j_{1}...j_{k}}$ transforms in the linear Êrepresentation of $SL(n)$
with highest weight
$\underline{\lambda}_{k}$. Thus
$ {\cal F}_{Sl(n)\otimes GL(1)\ \underline{i}_{1}...\underline{i}_{k}}$
transforms as a non-linear representation constructed from a linear
representation in the standard way (see appendix A, Êequation (A.8)), in
this case its non-linear transformations are contained in a matrix
belonging to $SO(n)$. This is consistent with the usual action of the
tangent space group on Êthe Êvielbein. Hence, if we use tangent space
internal indices then the $SL(n)$ symmetry will be essentially manifest
as long as we construct $SO(n)$ invariants.
\par
If we denote the part of the group element of the non-linear realisation
of $SL(n)$ with local subgroup $SO(n)$ which contains the Cartan
generators
${\underline H}$ of $SL(n)$ by $g_{Sl(n)}=e^{\underline{H}.\underline
{\phi}_{k}}$ then the
dimensionally reduced field strength $ {\cal
F}_{Sl(n)\otimes GL(1)\ \underline{i}_{1}...\underline{i}_{k}} $ carries
a factor of
$e^{\underline{\phi}.[{\underline Ê\lambda}_{k}]}$ where $\underline
\lambda_{k}$ is the
$SL(n)$ representation with highest weight $\underline{\lambda}_{k}$ and
$[\underline {\lambda}_{k}]$ is a weight in this representation. When
written in terms of the field strengths using equation (2.3) we also
find
exponential factors involving the dilaton. One can incorporate these
automatically by considering the group
$SL(n)\otimes GL(1)$ with the group element
$g_{Sl(n)\otimes GL(1)}=e^{\phi R}e^{\underline{H}.\underline
{\phi}_{k}}$ where $R$ is the generator of $GL(1,R)$.
\par
The dimensional reduction of terms containing field strengths in the IIA
higher derivative theory will Êlead to terms containing the object $
{\cal
F}_{Sl(n)\otimes GL(1)}$ Êof equation (2.10) multiplied by
exponentials of the field $\rho$ which arise from dimensional
reduction
using the ansatz of equation (2.6). ÊThe field strength of equation
(2.10) has a $\rho$ factor given by $e^{-k\beta\rho}$. ÊIf we were to
convert the space-time world indices to tangent indices then we would
also
acquire a factor $e^{-p\alpha\rho}$ if we have $p$ space-time indices.
\par
The derivatives of the scalars, including the dilaton, are contained
in the part of the Cartan form of $SL(n)\otimes GL(1)$ which changes
by a minus sign under the action of the Cartan involution,
and is denoted by $P_{SL(n)\otimes GL(1)}$. The action of the Cartan
involution on $SL(n)$ generators is such as to lead to $SO(n)$ being
the invariant
group
and on the generator $R$ it acts with a minus sign.
\par
After dimensional reduction, the IIA theory including the higher
derivative terms can be expressed in terms of the scalar curvature
$R$, which is an
$SL(n)$ singlet, the $P_{SL(n)\otimes GL(1,R)}$
ÊÊpart of the Cartan forms of $SL(n)\otimes GL(1,R)$
and the field strengths ${\cal F}_{SL(n) \otimes GL(1,R)\
\mu_{1}...\mu_{p} \underline i_{1} ... \underline i_{k}}$ which
transform
as non-linear representations of
$SL(n)\otimes GL(1,R)$.
\bigskip
\noindent
{\bf {3. The $E_{n+1}$ formulation in $d$ dimensions}}
\bigskip
As is well known, Êthe type II supergravity theories in $d$ dimensions
possess an $E_{n+1}$ symmetry [7-11,4]. Their actions are bilinear in
the
space-time derivatives and include the Riemann curvature, and
squares of the field strength and the derivatives of the scalars.
The metric, in Einstein frame, transforms as a singlet of
$E_{n+1}$ and therefore the Riemann curvature is invariant under
$E_{n+1}$ transformations.
The scalars belong to the non-linear realisation of $E_{n+1}$ with a
local subgroup $H_{n+1}$ which is the maximal compact subgroup. ÊThe
latter is just the Cartan involution invariant subgroup. ÊThis means
that the scalars are contained in a group element $g_{E_{n+1}} \in
E_{n+1}$ which transforms as $g_{E_{n+1}}\to g_0 g_{E_{n+1}}$ where
$g_0\in E_{n+1}$ is independent of space-time and also $g_{E_{n+1}}\to
g_{E_{n+1}}h$ where $h\in H_{n+1}$ and is an arbitrary function of
space-time. ÊÊWe can write the Cartan subalgebra part of
the group element as
$g_{E_{n+1}}= e^{\vec\phi.\vec{H}}$ where $\vec{H}$ are the $n+1$
Cartan
subalgebra generators of
$E_{n+1}$, which we have written as a vector. The
corresponding scalar fields are written as the vector $\vec\phi$.
\par
The non-linear realisation essentially specifies how the
scalars appear in the action. In particular, Êthe derivatives of the
scalars occur as Cartan forms of
$E_{n+1}$ in the coset directions. ÊIn terms of our group element
$g_{E_{n+1}}$, the Cartan forms which are given by $g_{E_{n+1}}^{-1}d
g_{E_{n+1}}$ in the coset directions, are denoted $P_{E_{n+1}}$. ÊWhen
evaluated they contain the roots
$\vec \alpha$ in Êthe form
$e^{\vec\phi.\vec\alpha}$ where $\vec\alpha$ are the roots of $E_{n+1}$.
\par
The gauge fields occur in the field strengths
$F$ that transform as Êlinear representations of $E_{n+1}$ with highest
weight $\vec{\Lambda}$ say. ÊHowever, we can convert a linear
representation of
$E_{n+1}$ into a non-linear representation using a group Êelement
$g_{E_{n+1}}^{-1}$. Explicitly, the non-linear representation
$|{\cal F} \rangle$ constructed from Êa linearly realised field strength
$| F\rangle$ is given by [32]
$$
| {\cal F}_{E_{n+1}} \rangle = L(g_{E_{n+1}}^{-1}) | F \rangle ,
\eqno(3.1)$$
where $L((g_{E_{n+1}}(\xi))^{-1})$ is the representation with highest
weight Ê$\vec{\Lambda}$.
ÊFrom equation (3.1) we find that the non-linearly
realised field strength $| {\cal F}_{E_{n+1}} \rangle $ contains a
dependence on the scalars $\vec\phi$ which is given by
$e^{\vec
\phi.[\vec{\Lambda}]}$ where $[\vec{\Lambda}]$ is a weight in
the
the $E_{n+1}$ representation with highest weight $\vec{\Lambda}$.
\par
If one dimensionally reduces, for example, the IIA Êsupergravity action
in ten dimensions and keeps track of the scalars that appear in the form
$e^{\vec\phi. \vec w}$ then one finds that $\vec w$ are
proportional to the roots of
$E_{n+1}$. Indeed, this is the simplest way to see that the
dimensionally
reduced theory is very likely to possess a $E_{n+1}$ symmetry.
\par
Using the same arguments, a generic higher derivative term in $d$
dimensions can be written as
a polynomial in the Riemann curvature, the non-linearly realised field
strengths and Cartan forms
$P_{E_{n+1}}$, but it is also multiplied by a function
of the scalar fields. Assuming that the higher derivative term as a
whole
is invariant under an $E_{n+1}$ transformation Êimplies that this
non-holomorphic function must be an object that transforms under
$E_{n+1}$ transformations like an automorphic form. ÊWe can expect that
this Êautomorphic form is built out of a particular representation, of
$E_{n+1}$ with highest weight $\vec{\Lambda}$ say. We write
the Êstates of this Êrepresentation in the form Ê$| \psi
\rangle=n_i|\vec{\mu}_i>$ where
$|\vec{\mu}_i>$ are a basis of the representation,
$\vec{\mu}_i$ are Êthe weights in the representation and $n_i$ are
integers.
To be more precise
it is constructed out of the non-linear representation Êof
$E_{n+1}$ constructed from this representation using the scalars, that
is, it is constructed out of the function $ |\varphi \rangle$,
defined by
$$
|\varphi \rangle Ê= ÊL(g_{E_{n+1}}^{-1}) | \psi \rangle.
\eqno(3.2)$$
It is obvious that $|\varphi \rangle$ Êcontains terms where the scalar
fields occur in the form
$e^{\vec\phi.[\vec{\Lambda}]}$. The automorphic Êform is a function of $
|\varphi \rangle $ and for the examples that are Êunderstood Êit
is of the generic form
$$
\sum_{n_i} <\varphi | \varphi >^{-s},
\eqno(3.3)$$
for some constant $s$. ÊThe automorphic form of equation (3.3) will
always contain a term with scalar field dependence given by $e^{-\sqrt
{2}s \vec{\Lambda}.\vec{\phi}}$, where $ \vec{\Lambda}$ is the
highest weight of the representation used to build the
automorphic form.
This construction is
described in more detail in reference [34]. The
use of integers corresponds to the fact that the symmetry group Êfor the
higher derivative terms is discretised since the charges of the theory
obey a quantisation condition. In fact we are using the Chevalley
definition of the discrete group; Êthat is the one generated by
$e^{\pm E_a}$,
$e^{\pm F_a}$ and $e^{\pm H_a}$ where $E_a$, $F_a$ and $H_a$ are the
Chevalley generators. ÊWe thank Lisa Carbone for this point.
\par
In this paper we will refer to the formulation of a higher derivative
term in $d$ dimensions Êjust described as
the $E_{n+1}$ formulation. A Êterm in the higher derivative
effective action will contain an Êexponential of the scalar
fields $\vec \phi$ of the form
$e^{\sqrt{2}
\vec w \cdot \vec \phi}$ where $\vec \phi$ is the field we introduced
earlier in ÊÊthis section.
Our task is to compare this with the equivalent factor that arises in
the dimensional reduction. However, in order to compare the
$E_{n+1}$ formulation of the type IIA theory in $d$ dimensions with the
dimensionally reduced formulation discussed in the previous section we
need to know the relationship between the fields that occur in the
dimensional reduction, namely the fields $ \underline \phi$, $\rho$ and
$\phi$, where $\underline \phi$ is an $n-1$-dimensional vector and those
that occur in the
$E_{n+1}$ formulation, Ênamely the
$n+1$-dimensional
$\vec\phi$. This will be given in the next section.

\bigskip
\noindent
{\bf {4. The ÊÊ$E_{11}$ formulation Ê}}
\bigskip
The eleven dimensional, IIA and IIB supergravity theories, as well
as the maximal type II supergravity theories in lower dimensions, can be
formulated as non-linear realisations [41,42,43]. ÊThe non-linear
realisations of the Kac-Moody algebra $E_{11}$, at low levels,
leads to all of these theories [42-47]. ÊAs such
$E_{11}$ encodes the fields of each of these theories and
provides us with a way of relating the fields in the different theories
to each other[55]. In fact the fields of these theories are in one to
one
correspondence with the generators of the Borel subalgebra of $E_{11}$
in the group decomposition, explained below, Êappropriate to each
theory. ÊIt has been conjectured that non-linear realisations of the
Kac-Moody algebra $E_{11}$ are extensions of all these supergravity
theories [42-47]. However, we will not use this conjectured $E_{11}$
result in this paper.
\par
A Kac-Moody algebra is formulated in terms of its Chevalley generators,
which include those in the Cartan subalgebra denoted by $H_{\hat a},
\ \hat a=1,2,\dots 11$. As such, the $E_{11}$ group element that occurs
in the non-linear realisation is of the form $g_{E_{11}}= e^{ \phi_{\hat
a} H_{\hat a}}$ provided we restrict our attention to the part that
is in
the Cartan subalgebra. Indeed, as there is an essentially unique
formulation of $E_{11}$ in terms of its Chevalley generators, Êthe
eleven
dimensional, IIA , IIB and $d$ dimensional theories viewed as
non-linear realisations have a common origin and their fields can be
mapped
into each other in a one to one manner [42-47]. It is this property
that we
are going to exploit.
\par
The $E_{11}$ Kac Moody algebra is encoded in the Dynkin
diagram
$$
\matrix {
& & & & & & & & 11 & & Ê&&\cr
& & & & & & & & \bullet & & && \cr
& & & & & & & & | & & &&\cr
\bullet & Ê- & Ê\ldots & -
&\bullet&-&\bullet&-&\bullet&-&\bullet&-&\bullet
\cr 1& & & & 6 & & 7 Ê& Ê& 8 & Ê& 9&&10 }
$$
\bigskip
\centerline {Figure 1. The $E_{11}$ dynkin diagram }
\bigskip

The {\bf eleven dimensional Êtheory} Êemerges if we decompose the
$E_{11}$ algebra in terms of the algebra that results from
deleting the exceptional node labelled eleven, namely the algebra
$GL(11)$. This subalgebra has the generators
$K^{\hat a}{}_{\hat b}$, $
\hat a, \hat b=1,\ldots ,11$ and it includes all the Cartan subalgebra
generators of $E_{11}$; the relation being [43]
$$
H_{\hat a}= K^{\hat a}{}_{\hat a}- K^{\hat a+1}{}_{\hat a+1},
\hat a=1,\ldots ,10,
$$
$$
H_{11}= - {1 \over 3}\left(K^{1}{}_{1}+...+ K^{8}{}_{8} \right) + {2
\over 3} \left( K^{9}{}_{9} + K^{10}{}_{10} + K^{11}{}_{11} \right).
\eqno(4.1)$$
The first ten generators being the Cartan subalgebra generators of
$SL(11)$.
\par
The contribution of the $GL(10)$ subgroup to the $E_{11}$ group
element in the
non-linear realisation is of the form
$$
e^{x^{\hat a}P_{\hat a}}e^{h_{\hat a}{}^{\hat Êb}K^{\hat a}{}_{\hat
b}},
\eqno(4.2)$$
where we have added the space-time translation
generators
$P^{\hat a}$. This is known to give rise to eleven dimensional gravity
and as a result the line in the above Dynkin diagram, Êthat is from
nodes
one to ten inclusive, Êis known as the gravity line. ÊIndeed the
Cartan form for this subgroup is given by
$$
g^{-1}dg= dx^{\hat \mu} \hat e_{\hat \mu}{}^{\hat a} P_{\hat a}
+(e^{-1}de)_{\hat a}{}^{\hat b} K^{\hat a}{}_{\hat b}.
\eqno(4.3)$$
It turns out that $e_\mu{}^a=(e^h)_{\hat{a}}{}^{\hat{b}}$ is the
eleven-dimensional vielbein.
\par
We may set the ÊÊdifferent formulations of the $E_{11}$ group element
to be equal and, restricting to the Cartan subalgebra, we find that
$$
e^{\hat \phi_{\hat a}\hat H_{\hat a}}= e^{h_{\hat a}{}^{\hat a} K^{\hat
a}{}_{\hat a}}.
\eqno(4.4)$$
Comparing coefficients of $K^{\hat
a}{}_{\hat a}$ using equation (4.1) we find the relations
$$
\hat{\phi}_{i} = h^{1}{}_{1} + h^{2}{}_{2} ... + Êh^{i}{}_{i} - {i
\over 2}
\sum_{j=1}^{11} h^{j}{}_{j}, \quad for \quad Ê1 \leq i \leq 8,
$$
$$
\hat{\phi}_{9}= h^{1}{}_{1} + h^{2}{}_{2} ... + Êh^{9}{}_{9} - 3 \sum_
{j=1}^
{11} h^{j}{}_{j},
$$
$$
\hat{\phi}_{10}= h^{1}{}_{1} + h^{2}{}_{2} ... + Êh^{10}{}_{10} - 2
\sum_
{j=1}^{11} h^{j}{}_{j},
$$
$$
\hat{\phi}_{11}= - {3 \over 2} \sum_{j=1}^{11} h^{j}{}_{j}.
\eqno(4.5)$$
The full non-linear
realisation of $E_{11}$ leads, at low levels and with the
decomposition to
$GL(11$), to the eleven dimensional supergravity theory. However, in
this
paper we are interested in only the fields associated with the Cartan
subalgebra parts of the algebra, hence the above restriction.
\par
Let us now consider the {\bf ten-dimensional ÊIIA theory} which is
obtained from eleven dimensions by dimensional reduction on a circle. In
this process, the diagonal components of the eleven dimensional metric
result in the diagonal components of the ten dimensional metric and a
scalar
$\phi$, which is the dilaton of the IIA theory.
\par
In terms of the $E_{11}$ formulation we obtain the IIA theory by
deleting nodes ten and eleven of the Dynkin diagram below (see figure
2) leaving us with a $GL(10)\otimes GL(1)$ algebra; the $GL
(10)$ algebra leads to ten
dimensional gravity, for the same reasons as occurred above in eleven
dimensions, and the $GL(1)$ factor leads to the IIA dilaton.
$$
\matrix {
& & & & & & & & 11 & & 10 \cr
& & & & & & & & \bullet & & \bullet \cr
& & & & & & & & | & & |\cr
\bullet & Ê- & Ê\ldots & - &\bullet&-&\bullet&-&\bullet&-&\bullet Ê\cr
1& & & & 6 & & 7 Ê& Ê& 8 & Ê& 9 }
$$
\medskip
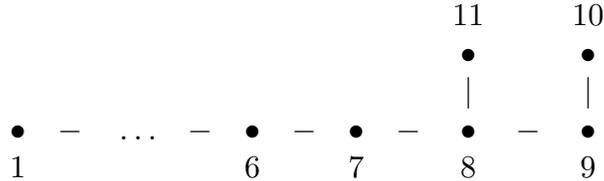
\centerline {Figure 2. The $E_{11}$ Dynkin diagram appropriate to the
IIA
theory }
\medskip
The gravity line is now the horizontal line of the Dynkin diagram of
figure 2. ÊÊÊThe IIA
supergravity theory emerges from the non-linear realisation of
$E_{11}$ with this decomposition.
\par
Let us denote the
generators of $GL(10)$ Êby
$ K^a{}_b,
\ \ a, b=1,\ldots , 10$ and let $R$ be the $GL(1)$ generator.
These contain the generators of the Cartan subalgebra of $E_{11}$. ÊThe
group element in the Cartan subalgebra of
$E_{11}$ can therefore be written in the form
$$
g=e^{\tilde h^{a}_{\ a} K^{a}_{\ a}}e^{\sigma R}
\eqno(4.6)$$
The tilde distinguishes the field from that in eleven
dimensions. However, in terms of the Chevalley generators in the Cartan
subalgebra of $E_{11}$, the group element has the same form as in eleven
dimensions, namely $g=e^{
\phi_{\hat a} H_{\hat a}}$.
\par
It turns out that the Cartan sub-algebra generators $ H_{\hat a}$
of the $E_{11}$ algebra and those in the $GL(10)\otimes GL(1)$ algebra
are related by [43]
$$
H_{a}= K^{a}_{\ a}- K^{a+1}_{\ a+1} , \quad
a=1,...,9,
$$
$$
H_{10}=-{1 \over 8}\left( K^{1}{}_{1}+...+ K^{9}{}_{ 9} \right) + {7
\over 8} K^{10}{}_{ 10} - { 3 \over 2 }R,
$$
$$
H_{11}=-{1 \over
4}\left( K^{1}{}_{ 1} +...+ K^{8}{}_{ 8} Ê\right) + {3\over 4}\left(
K^{9}{}_{ 9}+ K^{10}{}_{ 10} Ê\right) + R.
\eqno(4.7)$$
We note the useful relation $R={1\over 12}(-\sum_{a=1}^{10} K^a{}_a
+8K^{11}{}_{11})$.
\par
Equating the group element $g$ in the Cartan subalgebra written in
terms of the two different sets of generators we find that
$$
g= e^{\sum_{a=1}^{10}\tilde h^{a}_{\ a} K^{a}_{\ a}}
e^{\sigma R}
=e^{\phi_{1} \left( ÊK^{1}{}_{ 1}- K^{2}{}_{ 2}
\right)}
...e^{\phi_{9}\left( ÊK^{9}{}_{ 9} - ÊK^{10}{}_{ 10} \right)}
$$
$$
e^{\phi_{10}\left( - {1 \over 8}\left( ÊK^{1}{}_{ 1} +...+
K^{9}{}_{ 9} \right) + {7 \over 8} ÊK^{10}{}_{ 10} - {3 \over
2}R\right)}
e^{\phi_{11}\left( - {1 \over 4}\left( ÊK^{1}{}_{ 1} +...+
K^{8}{}_{ 8} \right) + {3 \over 4} \left( ÊK^{9}{}_{ 9} + ÊK^{10}{}_
{
10} \right) + R \right)}.
\eqno(4.8)$$
Comparing the coefficients of the generators $R$ and
$K^{a}{}_{a}$ Êwe find that
$$
\sigma=-{3 \over 2}\phi_{10} + \phi_{11},\quad
\tilde h^{1}{}_{ 1} = \phi_{1} - {1 \over 8}\phi_{10}
- {1 \over 4} \phi_{11} ,\quad
$$
$$
\tilde h^{i}{}_{ i} = -\phi_{i-1} + \phi_{i} - {1 \over 8}\phi_{10}
- {1 \over 4} \phi_{11},\quad ÊÊfor \ \ 2 \leq i<9 ,\quad
\tilde h^{9}{}_{ 9} = - \phi_{8} + \phi_{9} - { 1 \over 8 Ê}
\phi_{10} Ê+ {3 \over 4 } \phi_{11} ,\quad
$$
$$
\tilde h^{10}{}_{ 10} = - \phi_{9} + \phi_{10} + { 7 \over 8 Ê}
\phi_{10} Ê+ {3 \over 4 } \phi_{11}.
\eqno(4.9)$$
\par
We now consider the $E_{11}$ formulation of the {\bf $d$-dimensional
maximal supergravity theory} [42-47]. In the previous section we
dimensionally
reduced the IIA theory using the ansatz of equation (2.6) to find the
IIA
dilaton
$\phi$ of the original theory, and the $10-d$ scalars $\underline \phi$
arising from the diagonal components of the metric $G_{ij}$ and the
field
$\rho$.
\par
ÊFrom the $E_{11}$ perspective, the $d$-dimensional type II
supergravity
theory is found by writing the $E_{11}$ Dynkin diagram in the form
given in figure 3 below.
$$
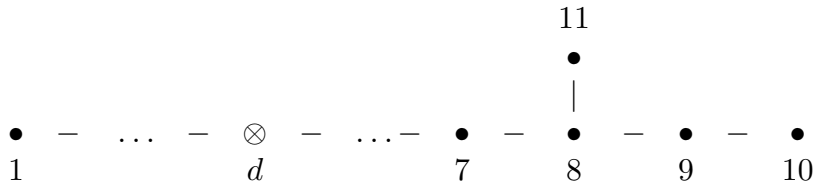
\matrix {
& & & & & & & & & 11 & & Ê&&\cr
& & & & & & & & Ê& \bullet & & && \cr
& & & & & & & & & | & & &&\cr
\bullet & Ê- & Ê\ldots & -
&\otimes&-&\ldots -&\bullet&-&\bullet&-&\bullet&-&\bullet
\cr 1& & & & d & & &7 Ê& Ê& 8 & Ê& 9&&10 }
$$
\medskip
\centerline {Figure 3. The $E_{11}$ Dynkin diagram appropriate to the}
\centerline {$d$-dimensional maximal supergravity Êtheory}
\medskip
Deleting node $d$ we find the residual algebra $E_{n+1}\otimes
GL(d)$; the latter algebra leads in the non-linear realisation Êto
$d$-dimensional gravity and the former algebra is the U-duality
group. ÊDecomposing the $E_{11}$ non-linear realisation into
representations of $E_{n+1}\otimes
GL(d)$ we find, at low levels, Êthe field content of the maximal
supergravity
theory in $d$ dimensions and indeed the supergravity theory itself.
We can further delete nodes ten and eleven, corresponding to the
dimensional
reduction of the IIA theory, and then we find the subalgebra $GL(d)
\otimes
SL(n)\otimes GL(1)\otimes GL(1)$, in other words the $E_{n+1}$ algebra
has been decomposed into $SL(n)\otimes GL(1)\otimes GL(1)$. The
$SL(n)$ arises from the line in the Dynkin diagram from nodes $d+1$ to 9
inclusive while the $GL(1)\otimes GL(1)$ factors
are essentially the Cartan subalgebra elements associated with nodes
ten and eleven. ÊSince the $SL(n)$ algebra is part of the $SL(10)$
algebra,
arising from the Êgravity line, of the ten dimensional IIA theory we
conclude that it is the
$SL(n)$ symmetry Êpreserved by dimensional reduction on the $n$ torus as
discussed in section two. ÊWe note that $SL(n)\otimes GL(1)\otimes GL
(1)$
contains all
the Cartan subalgebra elements of $E_{11}$ and that one of the $GL(1)$
factors is the $GL(1)$ symmetry of the IIA theory, discussed above
figure two. These ÊÊtwo $GL(1)$ Êfactors lead in the non-linear
realisation
to the fields
$\phi$ and
$\rho$.
\par
The dimensional reduction ansatz of equation (2.6) is implemented in
terms of $E_{11}$ by
rewriting the group element of the IIA theory of equation (4.6) in the
form
$$
g=e^{\dot h^{a}_{\ a} K^{a}_{\ a} + e_{1}\rho
\sum_{a=1}^{d}K^{a}_{\ a}}e^{\dot h^{i}_{\ i} K^{i}_{\ i} +e_{2}\rho
\sum_{i=d+1}^{10}K^{i}_{\ i}}e^{\sigma R}.
\eqno(4.10)$$
Here the $K^{a}{}_{ b}$, $a,b=1,\ldots d$, are the generators of the
$GL(d)$
algebra associated with $d$-dimensional gravity,
$K^{i}{}_{j},\ i,j=1,\ldots ,n$ are the generators of $SL(n)\otimes
GL(1)$ and $e_1$ and $e_2$ are constants. We have put a dot on the $h
$ fields to
distinguish them from the analogous fields used earlier in ten and
eleven dimensions. Taking into account the introduction of the field $
\rho$ we
set
$$
\dot Êh^{d+1}{}_{ d+1} + \dot h^{d+2}{}_{ d+2} +...+ \dot
h^{10}{}_{ 10} = 0.
\eqno(4.11)$$
\par
Introducing Êtranslation generators in $d$-dimensional
space-time and internal space
ÊÊinto the group element by including the factor
$e^{x^{a}P_{a}+x^{i}P_{i} }$ and computing the Cartan forms we find the
terms involving these new generators
are given by the expression
$g^{-1}\left(dx^{a}P_{a}+dx^{i}P_{i} \right)g$
which implies the identification
$$ \sum_{a=1}^{d}\left(
e^{\dot h^{a}_{\ a}+e_{1} \rho} \right)dx^{a}P_{a}+
\sum_{i=d+1}^{10}\left( e^{\dot h^{i}_{\ i}+e_{2} \rho} \right)dx^{i}
P_{i}
=
e^{\alpha \rho} e_{\mu}^{\ a}dx^{\mu}P_{a} Ê+ Êe^{\beta \rho} e_{j}^{\
i}dx^{j}P_{i}.
\eqno(4.12)$$
Taking $e_{1}= \alpha$, $e_{2}= \beta$ we do indeed recover the
vielbeins as they appear in the dimensional reduction ansatz of equation
(2.5) provided we identify $ Êe_\mu{}^a= (e^{\dot h})_\mu{}^a$ with the
vielbein in $d$-dimensional space-time and $ Êe_{i}{}^{\underline j}=
(e^{\dot h})_{i}{}^{\underline j}$ with the vielbein in the
$n$-dimensional internal space. Equation (4.11) implies that this latter
vielbein satisfies
the constraint $\det
e_i{}^{\underline j}=1$ as required in the
dimensional reduction ansatz. To find Êequation (4.12) we
have dropped Êvarious factors involving exponentials of the trace
of $h$ ÊÊas these are interpreted as
$\det e$.
\par
We now discuss the $E_{n+1}$ formulation of the $d$ dimensional
theory given in section three from the view point of the $E_{11}$
non-linear realisation. For simplicity we will consider only the case
$d\leq 7$. We saw from figure 3 that deleting node
$d$ leads to the algebra $GL(d)\otimes E_{n+1}$. By examining the $E_
{11}$
algebra one can find the generators of $GL(d)\otimes E_{n+1}$ in
terms of
those of $K^a{}_b$, $a,b=1,\ldots ,11$, $R^{a_1a_2a_3}$ etc.
One finds that the generators of
$GL(d)$ are $K^a{}_b$, $a,b=1,\ldots ,d$ and the Chevalley
generators Ê$T_a$, $a=d+1,\ldots , 11$ in the Cartan subalgebra
generators
of
$E_{n+1}$ Êare given by
$$
T_{d+1}=\dot K^{d+1}{}_{ d+1}-\dot K^{d+2}{}_{ d+2}, \ldots
T_{9}=\dot K^{9}{}_{ 9}-\dot K^{10}{}_{ 10},
$$
$$
T_{10}=-{1 \over 8}\left( \dot K^{d+1}{}_{ d+1} +...+ \dot K^{9}{}_
{9}
\right) + {7 Ê\over 8 } \dot K^{10}{}_{ 10}- {3 \over 2} \dot R,
$$
$$
T_{11}=-{1 Ê\over 4 } \left( \dot K^{d+1}{}_{ d+1} +...+ \dot K^{8}{}_
{ 8}
\right) + {3 \over 4} \left( Ê\dot K^{9}{}_{ 9} + \dot K^{10}_{\
\ 10}
\right) +\dot R.
\eqno(4.13)$$
where $\dot K^a{}_b= K^a{}_b -{1\over d-2}\delta ^a_b \sum_{c=1}^{d}
K^c{}_c$ for $a, b=d+1,\ldots , 10$ and $\dot R=R$. We note that Êthe
generators
$\dot K^a{}_b$ obey the necessary condition
$[\dot K^a{}_b,P_c]=0$ for $a,b=1,...,10$ and $c=1,\ldots, d$. In
this last equation
we have used the commutator $[K^a{}_b , P_c ]= -\delta^a_c P_b+{1
\over 2}
\delta^a_b P_c$. It
is straightforward to verify that
$T_a=H_a$, $a=d+1,\ldots , 11$ where these $H_a$ are the Chevalley
generators of $E_{11}$ given in equation (4.1), or equivalently from the
IIA viewpoint in Êequation (4.7). The $E_{11}$ group element
written in a way that displays the $GL(d)\otimes E_{n+1}$ decomposition
required in
$d$ dimensions and restricted to lie in Êthe Cartan subalgebra can be
written as
$$
g= e^{\sum_{a=1}^{d} \dot h^{a}_{\ a} K^{a}_{\ a}}
e^{\sum_{a=d+1}^{11} \varphi_a T_a}.
\eqno(4.14)$$
We have used that the $GL(d)$ generators are $K^a{}_b$, $a,b=1,
\ldots , d$
and denoted the $E_{n+1}$ fields by $\varphi_a$, $a=d+1,\ldots , 11$.
\par
We can now equate Êthe two different ways of expressing the $E_{11}$
group element given in equations (4.10) and (4.13), that is the one
that implements the dimensional reduction from the IIA theory to the one
that has the $GL(d)\otimes E_{n+1}$ decomposition in $d$ dimensions.
Using equations (4.13) and (4.7) Êand keeping only terms involving
$K^a{}_a$, $a=d+1,\ldots , 11$ we find the equation
$$
e^{ \left( \dot h^{d+1}{}_{ d+1}+e_{2}\rho \right) K^{d+1}{}_{
d+1} }...e^{ ÊÊ\left(\dot Êh^{9}{}_{ 9}+e_{2}\rho \right) K^{9}_{\
\ 9}
}e^{ \left(\dot Êh^{10}{}_{ 10}+e_{2}\rho \right) K^{10}{}_{ 10}
}e^{\sigma R}
$$
$$
= Êe^{\varphi_{d+1} \left(K^{d+1}{}_{ d+1}
-K^{d+2}_{\
\ d+2} \right) Ê}...e^{\varphi_{8} \left( K^{8}{}_{ 8}-K^{9}{}_{ 9}
\right) } e^{
\varphi_{9} \left( K^{9}{}_{ 9} - K^{10}{}_{ 10} \right) Ê}
$$
$$
e^{\varphi_{10}
\left( -{1 \over 8} \left( K^{d+1}{}_{ d+1} +...+ K^{9}{}_{ 9}
\right) + { 7 \over 8 }K^{10}{}_{ 10} -{ 3 \over 2 }R \right)}
e^{\varphi_{11} \left( -{1 Ê\over 4 } \left( K^{d+1}{}_{ d+1} +...+
K^{8}{}_{ 8} \right) + {3 \over 4} \left( ÊK^{9}{}_{ 9} + K^{10}_{\
\ 10} \right) + R \right)}.
\eqno(4.15)$$
Equating the coefficients of the generators $K^a{}_a$ and $R$
we
find the relations
$$
\eqalign{\dot h^{d+1}{}_{ d+1}+e_{2} \rho &= \varphi_{d+1} - {1
\over 8}
\varphi_{10} - {1 Ê\over 4 } \varphi_{11}, \cr
\dot h^{d+2}{}_{ d+2}+e_{2} \rho &= -\varphi_{d+1} + \varphi_{d+2}
- {1
\over
8}
\varphi_{10} - {1 Ê\over 4 } \varphi_{11}, \cr
... \cr
\dot h^{8}{}_{ 8}+e_{2} \rho &= -\varphi_{7} + \varphi_{8} - {1
\over 8}
\varphi_{10} - {1 Ê\over 4 } \varphi_{11}, \cr
\dot h^{9}{}_{ 9}+e_{2} \rho &= -\varphi_{8} + \varphi_{9} - {1
\over 8}
\varphi_{10} + {3 \over 4} \varphi_{11}, \cr
\dot h^{10}{}_{ 10}+e_{2} \rho &= -\varphi_{9} + { 7 \over 8 }\varphi_
{10}
+ {3
\over 4} \varphi_{11}, \cr
\sigma &= -{3 Ê\over 2 } \varphi_{10} + \varphi_{11}.}
\eqno(4.16)$$
\par
Solving these equations for the $E_{n+1}$ fields we find that
$$
\varphi_{i}= \dot h^{d+1}_{\ \ d+1} +\dot h^{d+2}_{\ \ d+2}
+ ... +
\dot h^{i}_{\
\ i}+ \left( n-10+i \right){ 8 \over 8-n } Êe_{2} \rho, \ \ d+1 \leq i <
8,
$$
$$
\varphi_{9} = \dot h^{d+1}{}_{d+1} + \dot h^{d+2}{}_{d+2}+ ... + \dot h^
{9}{}_{9} + {5n-8 \over 8-n} e_{2} \rho - {1 \over 4} \sigma,
$$
$$
\varphi_{10}=-{ 1 \over 2 } Ê\sigma + { 2 \over 8-n } n e_{2} \rho,
$$
$$
\varphi_{11}={1 Ê\over 4 } \sigma + { 3 \over 8 - n } n e_{2} \rho.
\eqno(4.17)$$

In this section we have formulated the $E_{11}$ algebra in terms
of the Chevalley generators, in particular the Cartan subalgebra
generators $H_a$, $a=1,\ldots ,11$, however, in section three we used
the Cartan-Weyl basis with generators $H_i, i=1,\ldots ,11$. ÊÊThe
advantage of the latter basis is that acting on a state $|\Lambda >$ of
weight $\Lambda_i$, the generators $H_i$, Êby definition, read off
the weight i.e. $H_i |\Lambda >=\Lambda_i Ê|\Lambda >$. ÊÊThe two
sets of
generators are related by $\alpha_a^i H_i= H_a$ where
$\alpha_a$ are the simple roots and $\alpha_{a}^{i}$ is the i'th
component. If we denote the fields in the ÊCartan-Weyl basis by
$\tilde{\varphi}_{a}$. The corresponding fields are related by
$\tilde \varphi^i H_i=\varphi^a H_a$ which implies the relation
$$
\tilde{\varphi}^{i}=\varphi^{a} \alpha_{a}^{i},
\eqno(4.18)$$
where $ \alpha_{a}$ are the simple roots of $E_{n+1}$ and the sum is
over
$a=d+1, \ldots , 11$ and the same for $i$. In addition, the fields
$\tilde{\varphi}_{i}$ in Êthe
$E_{11}$ group element
are equal to the fields $\varphi_{i}$ that appear Êin the automorphic
form,
up to a numerical factor. ÊWe see, through comparing the
normalisations of
the fields in the the $E_{11}$ group element $e^{\tilde{\varphi}_{i}
H_{i}}$
and the automorphic form group element $e^{-{1 \over \sqrt{2} }
\vec{\varphi}.\vec{H}}$, that
$$
\vec{\varphi} = \left(-\sqrt{2} \tilde{\varphi_{1}}, - \sqrt{2}\tilde
{\varphi_
{2}}
, - Ê\sqrt{2}\underline{\tilde{\varphi}} \right)
\eqno(4.19)$$
Using equation (B.3) of appendix B Êand equation (4.18) we then find
that
the components of the $E_{11}$ group element fields $\tilde{\varphi}_{i}
$ in
Cartan-Weyl basis and the Chevalley basis are related by
$$
\tilde{\varphi}^{1}=x\varphi^{10}
=x\left(-{ 1 \over 2 } \sigma+\left({ 2 \over 8-n } \right) n e_{2} \rho
\right),
\eqno(4.20)$$
$$
\eqalign{\tilde{\varphi}^{2}&=-{
\underline{\lambda}_{n-2}.\underline{\lambda}_{n-1} \over y} \varphi_
{10} +
y\varphi_{11} \cr
&=-{ \underline{\lambda}_{n-2}.\underline{\lambda}_{n-1} \over y}\left
(-{
1 \over 2 } \sigma+\left({ 2 \over 8-n } \right) n e_{2} \rho \right) Ê+
y\left({1 Ê\over 4 }\sigma+\left({ 3 \over 8 - n } \right) n e_{2} \rho
\right),}
\eqno(4.21)$$
and
$$
\underline{\tilde{\varphi}} = Ê\sum_{i=d+1}^{9} \varphi_{i}\underline
{\alpha}_{i-d} - \varphi_{10} \underline{\lambda}_{n-1} - \varphi_{11}
\underline{\lambda}_{n-2}.
\eqno(4.22)
$$
Note that
$$
\underline{\tilde{\varphi}}.\underline{\alpha}_{i}=\dot h^{i}_{\ \
i}- \dot h^{i+1}_{\ \ i+1}, \quad
\underline{\tilde{\varphi}}.\underline{\lambda}_{j}=\sum_{i=d+1}^{d+j}
\dot h^{i}{}_{i}. \eqno(4.22)$$

\bigskip
\noindent
{\bf {5. Constraints on the automorphic forms}}
\bigskip
In this section we will compare the $E_{n+1}$ formulation in $d$
dimensions given in section three with the results of section two
found by dimensionally reducing the IIA theory in ten dimensions; as a
result we will Êfind constraints on the automorphic forms. In order to
carry out the comparison we will use the field relations of the last
section. ÊThe field strengths occurred in the $E_{n+1}$ formulation
as the
non-linear representations ${\cal F}_{E_{n+1}}$ Êgiven in equation
(3.1),
while the derivatives of the scalars occur in the Cartan forms
$P_{E_{n+1}}$. These are constructed using the group element
$g_{E_{n+1}}$, however, this Êis just the
$E_{11}$ group element restricted to lie in the subalgebra $E_{n+1}$
and it is given below equation (4.12). We noted that if one deletes
nodes 10 and 11 in the $E_{11}$ Dynkin diagram the $E_{n+1}$ algebra
is reduced to
$SL(n) \times GL(1) \times GL(1)$.
\par
In the dimensional reduction of the IIA theory we found a manifest
$SL(n)\otimes GL(1)$ symmetry; the first factor arises from
the diffeomorphisms preserved by the torus while the second factor
is the $GL(1)$ symmetry of the IIA theory in ten dimensions. As such,
the
field strengths that appear in the dimensional reduction can be
expressed in terms of the non-linear representation Êof $SL(n)\otimes
GL(1)$ denoted by ${\cal F}_{Sl(n)\otimes GL(1)}$ and the derivatives
of the scalars in
terms of the Cartan forms $P_{sl(n)\otimes GL(1)}$.
\par
Deleting nodes ten and eleven of the Dynkin diagram of figure 3 we find
that $E_{n+1}$ decomposes into $SL(n)\otimes GL(1)\otimes GL(1)$ and
one can carry out the decomposition of the non-linear representations
that
occur in the $E_{n+1}$ formulation.
Clearly, the non-linear representations of the field strengths ${\cal
F}_{E_{n+1}}$ will decompose into the non-linear representations ${\cal
F}_{sl(n)\otimes GL(1)}$ with appropriate factors corresponding to the
additional $GL(1)$. ÊThe same discussion applies to the derivative of
the
scalars which appear in
$P_{E_{n+1}}$ and
$P_{SL(n)\otimes GL(1)}$. Given a particular term in the higher
derivative effective action found by dimensional reduction and matching
it with the $E_{n+1}$ formulation, Êthe
$SL(n)\otimes GL(1)$ parts will automatically agree and it is with the
comparison of the other
$GL(1)$ factor that we find non-trivial results.
\par
It would be instructive to systematically carry out the decomposition of
$E_{n+1}$ formulation when
decomposed to Ê$SL(n) \times GL(1) \times GL(1)$, but for our present
purposes it suffices to carry it out for the generators that belong to
the Cartan subalgebra. With this restriction the group element of
$E_{n+1}$ is given, below equation (4.12), Êby $g_{E_{n+1}}=
e^{H_a \varphi_a}$, ÊÊbut an equivalent ÊÊformulation, Êin terms of
the field variables associated with dimensional reduction, is given Êin
equation (4.15). ÊÊMatching these Êwe found in equations (4.16) and
(4.17)
how the fields
$\varphi_a,\ a=d+1,\ldots , 11$ correspond to the fields $\dot
h^a{}_a, a=d+1, \ldots, 10$, $\rho$ and $\phi$ found in the
dimensional reduction of the IIA theory. The additional $GL(1) \otimes
GL(1)$ group found in the reduction
then corresponds to the Cartan subalgebra generators $H_{10}$ and
$H_{11}$ or from the dimensional reduction view point to Êthe fields
$\rho$ and
$\phi$.
\par
We now consider the decomposition in more detail.
One may write any root of
$E_{n+1}$ in terms of its simple roots:
$$
\vec \alpha=m_{c}\vec \alpha_{n+1} + n_{c}\vec \alpha_{n} +
\sum_{i=1}^{n-1} m_{i} \vec \alpha_{i} = n_{c}\left(x,-{
\underline{\lambda_{n-2}}.\underline{\lambda_{n-1}} \over y} Ê,
\underline{0} \right) + m_c \left(0,y, \underline 0 \right) - \vec
\lambda,
\eqno(5.1)$$
where $\vec \lambda = m_{c} \lambda_{n-2} + n_{c} \lambda_{n-1} -
\sum_{i=1}^{n-1}k_{i} \underline{\vec \alpha_{i}} $ and we have used
equation (B.3). The roots of $E_{n+1}$ are labelled by the
integers $m_{c}, n_{c}$ which are referred to as the levels.
If a representation of $SL(n)$ occurs in the decomposition of the
adjoint representation of $E_{n+1}$ then its highest weight must
appear on
the right-hand side as one of the $\underline \lambda$'s. We can
examine which representations occur level
by level.
At level $n_c=m_c=0$ one obviously finds the
adjoint representation of $SL(n)$. At higher levels the
highest weights, and so representations, of $SL(n)$ that occur are
given in the table below
$$
\matrix{
m_c=1, \ n_c=0 & m_c=0, \ n_c=1 & m_c=1, \ n_c=1 & m_c=2, \ n_c=1 \cr
\underline \lambda_{2} & Ê\underline \lambda_{1}&
\underline\lambda_{3} & \underline 0 Ê\cr
\cr
m_c=3, \ n_c=1 & m_c=2, \ n_c=2 &m_c=3, \ n_c=2 & Ê\cr
\underline 0 Ê& Ê\underline \lambda_{6} Ê& \underline \lambda_{1} .}
\eqno(5.2)
$$
As such one finds that the weights in the adjoint representation of
$E_{n+1}$ are given by
$$
\left( 0, 0, \left[\underline \alpha _{1} + ... + \underline \alpha_
{n-1}
\right]
\right)
, Ê\left( 0 , y, \left[
\underline \lambda_2 Ê\right] \right), \left( x, -{
\underline{\lambda_{n-2}}.\underline{\lambda_{n-1}} \over y} , \left[
\underline \lambda_1 Ê\right] \right) ,
$$
$$
\left( x , -{
\underline{\lambda_{n-2}}.\underline{\lambda_{n-1}} \over y} Ê+ y,
\left[ \underline \lambda_3 Ê\right] \right)
, \left( Êx , -{
\underline{\lambda_{n-2}}.\underline{\lambda_{n-1}} \over y} Ê+ 2 y,
\underline 0 ÊÊ\right),
$$
$$
\left( Êx , -{
\underline{\lambda_{n-2}}.\underline{\lambda_{n-1}} \over y} Ê+ 3 y,
\underline 0 ÊÊ\right)
, \left( Ê2x , -2{
\underline{\lambda_{n-2}}.\underline{\lambda_{n-1}} \over y} Ê+ 2 y,
\left[ \underline \lambda_6 Ê\right] \right),
$$
$$
\left( Ê2x , -2{
\underline{\lambda_{n-2}}.\underline{\lambda_{n-1}} \over y} Ê+ 3 y,
\left[ \underline \lambda_1 Ê\right] \right).
\eqno(5.3)
$$
The Cartan form $P_{E_{n+1}}$ belongs to the adjoint representation of
$E_{n+1}$ and at level $m_c=n_c=0$ decomposes into the Cartan forms of
$SL(n)$. Using the
decomposition of equation (5.3) we see that at higher levels they
decompose as follows
$$
\matrix{
m_c=1, \ n_c=0 & m_c=0, \ n_c=1 & m_c=1, \ n_c=1 & m_c=2, \ n_c=1 \cr
P_{SL(n)i_{1}i_{2}}& ÊP_{SL(n)i} & P_{SL(n)i_{1}i_{2}i_{3}} Ê&
P_{SL(n)i_{1}i_{2}...i_{n}} \cr
\cr
m_c=3, \ n_c=1 & Êm_c=2, \ n_c=2 &m_c=3, \ n_c=2 & \cr
P_{SL(n)i_{1}i_{2}...i_{n}}& ÊP_{SL(n)i_{1}...i_{6}} & P_{SL(n)i}.
}
\eqno(5.4)$$
We noted previously that the Cartan form $P_{E_{n+1}}$ contains a
dependence on the scalars $\vec \phi$ in
the form Êfactor $e^{{1\over \sqrt{2} }\vec \phi .\vec \alpha}$.
Under the decomposition we find Êthe $SL(n)\otimes GL(1)$ Cartan
forms $P_{SL(n)\otimes GL(1)}$ and exponentials in $\rho$. ÊUsing
equations
(4.20)-(4.22), (B.3) and (B.4) we find that the latter factors at level
$m_c$,
$n_{c}$ are
$$
e^{\left(2m_c
+n_c\right)\alpha \rho \left({ 8-n \over n}\right)}.
\eqno(5.5)$$
\par
We now consider the terms that result from the dimensional reduction
from the IIA theory using the discussion of section two.
The ten dimensional origins of the decomposition of
the adjoint representation of $E_{n+1}$ at each level may be found by
examining the $SL(n)$ and
space-time index structure. In particular,we see that the
$SL(n)\otimes GL(1)$ Cartan forms $P_{SL(n)\otimes GL(1)}$ Êat levels
$\left(m_c=0, \ n_c=1\right)$,
$\left(m_c=1,
\ n_c=0\right)$
and
$\left(m_c=1, \ n_c=1\right)$ come from the dimensional reduction of the
two form field strength $\tilde {\cal F}_{a \underline i_1
\underline i_2}$, three form field strength $\tilde {\cal F}_{a
\underline
i_1
\underline i_2 \underline i_3}$ Êand four form field
strength $\tilde {\cal F}_{a \underline i_1 \underline i_2 \underline
i_3
\underline i_4}$ respectively. ÊThe ÊCartan forms, at higher
levels, are associated with the dimensional reduction of the dualised
two, three and four form fields strengths for levels $ \left( m_c=3, \
n_c=1 \right) $,
$ \left( m_c=2, \ n_c=2 \right) $ and $ \left( m_c=2, \ n_c=1 \right) $
respectively, along with the dualised graviphoton at level
$\left(m_c=3, \ n_c=2 \right)$. ÊWe note that the dualised four form
only
appears as a Cartan form of $SL(n)\otimes GL(1)$ in $d=5$, while the
dualised three form is only present as a Cartan form of $SL(n)\otimes
GL(1)$ in
$d=4$. ÊWhile the dualised graviphoton is a Cartan form of
$SL(n)\otimes GL(1)$ only in
$d=3$ and Êwe also find the dualised two form is
also realised as a Cartan form of $SL(n)$.
The Cartan forms of
$SL(n)\otimes GL(1)$, arising upon dimensional reduction, carry one
$d$ dimensional space-time index and $\left( 2m_c + n_c \right)$
internal
indices. ÊTherefore, each Cartan form of $SL(n)\otimes GL(1)$, at a
given
level, occurs with an exponential of $\rho$ which is given by
$$
e^{-\rho\left(\alpha +\left(2m_c+n_c \right)\beta
\right)}=e^{\left(2m_c+n_c\right) \alpha \rho \left({ 8-n Ê\over n }
\right)}e^{-\alpha \rho}.
\eqno(5.6)$$
Comparing with the result, given in equation
(5.5), of the $E_{n+1}$ formulation we find a surplus factor of
$e^{-\alpha \rho}$ multiplying the dimensionally reduced term.
We note that the factors involving $\phi$ and $\underline \phi$ will
match automatically due to the automatic agreement of the $SL(n) \times
GL(1)$ part.
\par
To treat the other building blocks in the same way we must learn how
to decompose
more general representations of $E_{n+1}$ into those of $SL(n) \times
GL(1) \times GL(1)$. To do this we use the technique of reference
[48]. If one wants to consider the representation of $E_{n+1}$ with
highest weight $\Lambda_{i}$, associated with the node labeled $i$,
we add a new node, denoted $\star$, to the $E_{n+1}$ Dynkin diagram
which is connected to the node labeled $i$ by a single line to
construct the Dynkin diagram for an enlarged algebra of rank $n + 2$.
Deleting the $\star$-node we recover the $E_{n+1}$ Dynkin diagram and
the representation of $E_{n+1}$ with highest weight $\Lambda_{i}$ is
found in the adjoint representation of the enlarged algebra provided
we keep only contributions at level $n_{\star} = 1$. Thus we find the
decomposition of the representation of $E_{n+1}$ with highest weight $
\Lambda_{i}$ into representations of $SL(n) \times GL(1) \times GL(1)
$ by decomposing the adjoint representation of the enlarged algebra
but deleting the additional node and keeping only contributions with
$n_{\star} = 1$ and deleting nodes $10$ and $11$
but keeping all levels of $m_{c}$ and $n_{c}$.
\par
In the $E_{n+1}$ formulation of the effective action in $d$ dimensions,
the one form gauge field, out of which the two form field strengths are
constructed, appear in the representation with highest weight $\vec
\Lambda_{1}$. ÊThe $\vec \Lambda_{1}$ representation of $E_{n+1}$ may be
decomposed into representations of $SL(n)$, with an associated type IIA
dilaton weight, level by level. ÊAt level $\left( m_{c} , n_{c}
\right) $
one finds
$$
\matrix{
m_c=0, \ n_c=0 & m_c=1, \ n_c=0 & m_c=0, \ n_c=1 & m_c=1, \ n_c=1 \cr
\underline \lambda_{1} & Ê\underline \lambda_{n-1} &
\underline 0 & ÊÊ\underline \lambda_{n-2} Ê\cr
\cr
m_c=2, \ n_c=1 & m_c=2, \ n_c=2 & m_c=3, \ n_c=1 & m_c=3, \ n_c=2 Ê\cr
\underline \lambda_{n-4} Ê& Ê\underline \lambda_{n-5} & Ê\underline
\lambda_{n-6} & ÊÊ\underline \lambda_{n-1}.
}
\eqno(5.7)$$
Therefore, the weights of the $\vec \Lambda_{1}$ representation are
$$
\left( { 1 \over 2x }, { \underline \lambda_{1}. \underline
\lambda_{n-2} \over y }, \left[\underline \lambda_{1} \right]
\right), \left( { 1 \over 2x } , { \underline \lambda_{1}. \underline
\lambda_{n-2} \over y }-y, \left[ \underline \lambda_{n-1} Ê\right]
\right), \left( { 1 \over 2x }-x , Ê{ \underline \lambda_{1}.
\underline \lambda_{n-2} \over y }+{ \underline \lambda_{n-1}.
\underline
\lambda_{n-2} \over y }, \left[ \underline 0 Ê\right] \right),
$$
$$
\left({ 1 \over 2x }-x, { \underline \lambda_{1}. \underline
\lambda_{n-2} \over y }-y+{ \underline \lambda_{n-1}. \underline
\lambda_{n-2} \over y } , \left[ \underline \lambda_{n-2} Ê\right]
\right),
$$
$$
\left( Ê{ 1 \over 2x }-x , { \underline \lambda_{1}.
\underline \lambda_{n-2} \over y }-2y+{ \underline \lambda_{n-1}.
\underline \lambda_{n-2} \over y } , \left[ \underline \lambda_{n-4}
\right] Ê\right),
$$
$$
\left( { 1 \over 2x }-2x , Ê{ \underline \lambda_{1}.
\underline
\lambda_{n-2} \over y }-2y+2{ \underline \lambda_{n-1}. \underline
\lambda_{n-2} \over y } , \left[ \underline \lambda_{n-5} \right]
\right),
$$
$$
\left( { 1 \over 2x }-x , Ê{ \underline \lambda_{1}.
\underline \lambda_{n-2} \over y }-3y+{ \underline \lambda_{n-1}.
\underline \lambda_{n-2} \over y } , \left[ \underline \lambda_{n-6}
\right] \right),
$$
$$
\left( { 1 \over 2x }-2x, Ê{ \underline \lambda_{1}. \underline
\lambda_{n-2} \over y }-3y+2{ \underline \lambda_{n-1}. \underline
\lambda_{n-2} \over y } , \left[ \underline \lambda_{n-1} Ê\right]
\right).
\eqno(5.8)
$$

ÊFrom the weights, we see that the Êcorresponding two form field
strengths, at each level, are
$$
\matrix{
m_c=0, \ n_c=0 & m_c=1, \ n_c=0 & m_c=0, \ n_c=1 & m_c=1, \ n_c=1 \cr
{\cal F}^i_{a_1 a_2}& {\cal F}_{a_{1} a_{2} i} Ê& {\cal F}_{a_{1} a_
{2} } & Ê{\cal F}_{a_
{1} a_{2} i_{1} i_{2}}
\cr
\cr
m_c=2, \ n_c=1 & m_c=2, \ n_c=2 & m_c=3, \ n_c=1 & m_c=3, \ n_c=2 Ê\cr
{\cal F}_{a_{1} a_{2} i_{1} ... i_4 } & Ê{\cal F}_{a_{1} a_{2}
i_1 ... i_5 } &
{\cal F}_{a_{1} a_{2} i_1 ... i_{n}} & ÊÊ{\cal F}_{a_{1} a_{2} i}.
}
\eqno(5.9)$$

After dualisation, any two form field Êstrength will appear as a one
form field strength in $d=3$ dimensions therefore we only need to
consider
two form field strengths in $d\geq4$ dimensions. ÊOne finds the
maximum level that contributes
is $\left(m_c=3, \ n_c=2\right)$, in the remaining dimensions any level
$\left( m_{c},n_{c} \right)$ listed in the above decomposition will
appear in $d$ dimensions if $\left( 2m_{c}+n_{c} -1 \right)\leq n $.
The two form field strength at level $\left(m_c, \ n_{c}\right)$ arises
through the dimensional reduction of the metric at level
$\left(0,0\right)$, three form field strength at level
$\left(1,0\right)$, two form field strength at level $\left(0,1\right)$
and four form field strength at level $\left(1,1\right)$. ÊThe higher
levels in the decomposition of the representation with highest weight
$\vec \Lambda_{1}$ are associated
with
the dimensional reduction of the dualised field strengths and the
graviphoton.
\par
A two form field strength in some representation of $SL(n)$ at level
$\left(m_c, ÊÊn_{c}\right)$ in the $E_{n+1}$ formulation of the IIA
theory appears multiplied by the factor
$$
e^{-{ 8 \over n}\alpha \rho-\left(2m_c
+n_c\right)\left({ 8-n Ê\over n }\right)\alpha \rho},
\eqno(5.10)
$$
where the factors associated with $SL(n)$ fields $\underline{\phi}$
and the
IIA dilaton $\phi$ match those found upon dimensional reduction.
Comparing the volume with the dimensionally reduced two form field
strengths, which carry two $d$ dimensional indices and $2m_{c}+n_c -1$
internal indices and as a result appear multiplied by the factor
$$
e^{-\rho\left(2 \alpha + \left( 2m_{c} +n_{c} -1 Ê\right) \beta
\right)}
= e^{-\alpha \rho} e^{-{ 8 \over n} -\left(2m_c + n_c\right) \left({ 8-n
\over n }\right) \alpha \rho },
\eqno(5.11)
$$
we find that the two form field strengths in the dimensionally reduced
type IIA effective action carry an additional factor of $e^{-\alpha
\rho}$.
\par
Three form field strengths appear in the type IIA effective action in $d
\geq 6$ dimensions. ÊIn the $E_{n+1}$ formulation, the two form gauge
fields, from which the three form field strengths are constructed,
lie in
the representation with highest weight $\vec \Lambda_{n}$. ÊThe $\vec
\Lambda_{n}$ representation decomposes into representations of $SL(n)$
with an associated type IIA dilaton weight, at level $\left( m_c, \ n_c
\right)$, in the following way
$$
\matrix{
m_c=0, \ n_c=0 & m_c=0, \ n_c=1 & m_c=1, \ n_c=1 & m_c=1, \ n_c=2 \cr
\underline{0} & Ê\underline{\lambda}_{n-1} &
\underline{ \lambda}_{n-3} & ÊÊ\underline{0} Ê.
}
\eqno(5.12)$$
This decomposition leads one to observe that the weights in the $\vec
\Lambda_{n+1}$ representation of $E_{n+1}$ are
$$
\left( { 1 \over x}, 0, \ \underline 0 Ê\right), \left(
{ 1 \over x}-x , \ { \underline{\lambda}_{n-2}.\underline{\lambda}_{n-1}
\over y}, \left[ \underline \lambda_{n-1} Ê\right] \right), \left(
{ 1 \over x}-x , \ { \underline{\lambda}_{n-2}.\underline{\lambda}_{n-1}
\over y} -y , \left[ \underline \lambda_{n-3} Ê\right] \right),
$$
$$
\left( { 1 \over x}-2x , \ { 2 \underline{\lambda}_{n-2}.
\underline{\lambda}_{n-1} \over y} -y Ê, \ \underline 0 \right).
\eqno(5.13)$$
The three form field strengths, at level $\left(m_c, \ n_c \right)$, are
$$
\matrix{
m_c=0, \ n_c=0 & m_c=0, \ n_c=1 & m_c=1, \ n_c=1 & m_c=1, \ n_c=2 \cr
{\cal F}_{a_1 a_2 a_3}& {\cal F}_{a_{1} a_{2} a_{3} i} Ê& {\cal F}_{a_
{1} a_{2} i_{1}
i_{2} i_{3} } & Ê{\cal F}_{a_{1} a_{2} a_{3} }.
}
\eqno(5.14)
$$
Any three form may be dualised to a lower degree form in $d \leq 5$,
therefore we need only consider three form field strengths for $n
\leq 4$.
For $n=4$ all of the three form field strengths listed above are
present.
For $n < 4$ a three form field strength, at level $\left(m_c, \ n_c
\right)$, will be present if $\left( 2m_c +n_c \leq n \right)$. ÊThe
origin of the three form field strengths is clear, the three form field
strength at level $\left(0, \ 0 \right)$ is the dimensionally reduced
three form field strength, while the three form field strength at level
$\left(0, \ 1 \right)$ is the dimensionally reduced four form field
strength. ÊThe remaining two levels are associated with the duals of the
dimensionally reduced three and four form field strengths. ÊThe
decomposition of the $\vec \Lambda_{n} $ of $E_{n+1}$, at level
$\left(m_c, \ n_c \right)$, leads to the $E_{n+1}$ formulation of the
non-linearly realised three form field strengths containing the
factor of
$$
e^{\left(-2 + \left(2m_c
+n_c\right)\left({ 8-n Ê\over n }\right) \right)\alpha \rho},
\eqno(5.15)$$
again, we find the factors involving the IIA dilaton $\phi$ and the
$SL(n)$ fields $\underline{\phi}$ agree with the dimensionally reduced
formulation. ÊHowever, the three form field strengths in the
dimensionally reduced formulation come with three space-time indices and
$2m_c +n_c$ internal indices, therefore they carry a factor of
$$
e^{-\rho\left(3 \alpha + \left( 2m_{c} +n_{c} Ê\right) \beta Ê\right)} =
e^{-\alpha \rho} e^{\left(-2 +\left(2m_c + n_c\right) \left({ 8-n Ê\over
n }\right)\right)\alpha \rho}.
\eqno(5.16)
$$

Comparing the $\rho$ factor of the $E_{n+1}$ formulation and the
dimensionally reduced formulation, one finds that the three form field
strengths in the dimensionally reduced effective action of the type IIA
theory carry an additional factor of $e^{-\alpha \rho}$. ÊThe four form
field strengths, which only exist in $d\geq 8 $
space-time dimensions follow the same pattern, with the dimensionally
reduced formulation containing an additional factor of $e^{- \alpha
\rho}$ when compared to the $E_{n+1}$ formulation of the effective
action
in $d$ dimensions.
\par
Thus, one finds that the surplus weight of any
derivative of the scalars
form or field strengths in the dimensionally reduced formulation of the
effective action of the type IIA theory in $d$ dimensions contains an
additional factor of $e^{- \alpha \rho}$ when compared to the $E_{n+1}$
formulation in $d$ dimensions. Thus we find an excess factor of
$e^{- \alpha \rho}$ for every space-time derivative in the effective
action. The
dimensionally reduced theory also carries a factor of $e^{-\tilde{s}
\phi}$ Êfrom the ten dimensional automorphic form, where $\tilde{s}$ is
given in equation (2.9) and is fixed by demanding that, upon
transforming
to string frame, any term carries a factor of $e^{\phi(2g-2)}$ arising
from a perturbative expansion in the ten dimensional IIA string coupling
constant
$g_{s}=e^{\phi}$ at order $g$. Also from the dimensional reduction of
the
$\det (e)$ from ten dimensions we find a factor of Ê$e^{- 2\alpha
\rho}$.
Therefore, we
find that the dimensionally reduced theory, when packaged up into
objects
transforming under
$E_{n+1}$, has Êa surplus factor of
$$
e^{ - \left( l_{T}-2 \right) \alpha \rho - \tilde{s} \phi Ê}.
\eqno(5.17)$$
where Ê$l_{T}=\tilde{l}_{0}+ \tilde{l}_{R}+
\tilde{l}_{1}+\tilde{l}_{2}+\tilde{l}_{3}+\tilde{l}_{4}$ is the total
number of derivatives and $\tilde{s}$ is given in equation (2.9).
We note that the factor of equation (5.17) can be written as
$e^{-\sqrt 2 \vec \Lambda_\phi \cdot \vec \varphi}$ Êwhere
$$
\vec \Lambda_{\phi} = \left( { \tilde{s} \over \sqrt{2} }, \alpha
{\left(
l_{T} - 2 Ê\right) \over \sqrt{2} }, \underline 0 Ê\right) = \left
({l_T -
2 \over 4 } + {3 \over 4}\left(l_{RR} -2g Ê\right) \right) \vec
{\Lambda}_{n} + Ê{1 \over 2 }\left(l_{RR} - 2g ÊÊ\right)\vec{\Lambda}_
{n+1}.
\eqno(5.18)$$
where $l_{RR}=\tilde{l}_{2}+\tilde{l}_{4}$ is the number of R-R fields
in a given term.
\par
The theory in $d$ dimensions contains an Ê$E_{n+1}$
automorphic form and this must account for the missing factors.
Therefore, we conclude that the automorphic form must contain the
weight $\vec \Lambda_{\phi}$.
For a pure NS-NS term at $g=0$, (i.e. setting $\tilde{l}_{2}$=$\tilde
{l}_{4}=0$ and $g=0$)
the leading order contribution to the automorphic form carries the
weight
$$
\vec \Lambda_{\phi} Ê= \left({l_T - 2 \over 4 } \right) \vec{\Lambda}_
{n} .
\eqno(5.19)$$
As such it is likely that the automorphic form is constructed
from the representation with Êhighest weight
$\vec \Lambda_{n}$ and with $s=\left({l_T - 2 \over 4 }
\right)$. ÊThe R-R terms are related to those in the NS-NS
sector by an $SL(2, R)$ rotation and so are automatically accounted
for.

\medskip
{\bf 6. Discussion of results and their consequences for the
automorphic forms of string theory}
\medskip
In this paper we have carried out the dimensional reduction Êof the
higher derivative corrections of the IIA theory and found that the
$E_{n+1}$ automorphic forms that appear as coefficients of the terms
in the
effective action in
$d=10-n$ dimensions must contain the fundamental weight
$\Lambda_n$ associated with Ênode $n$ of the $E_{n+1}$ Dynkin diagram of
figure 5; this corresponds to node ten in the $E_{11}$
Dynkin diagram of figure 3. ÊThe well understood $E_{n+1}$
automorphic forms that appear
in ÊÊstring theory Êare constructed using a
given representation of
$E_{n+1}$; the reader may, for example, Êconsult the explicit
construction
of these objects Êgiven in [34]. ÊÊAs such, the result of this paper
strongly suggests that Êthe automorphic forms that occur in string
theory Êare Êconstructed from the representation with highest weight
$\Lambda_n$. ÊÊMore precisely, Êit implies that if the coefficient of
the
higher derivative term is a sum of automorphic forms then one of them
should Êbe constructed Êfrom the highest weight
$\Lambda_n $ as it could happen that the other automorphic forms Êdo not
occur in the dimensional reduction from the IIA theory in ten
dimensions. ÊA similar Êanalysis from the IIB perspective gave the Êsame
result namely that the automorphic form contains the weight
$\Lambda_n$ [39]. However, from the M theory
perspective, that is from eleven dimensions, a similar ÊÊanalysis
found that the automorphic form contains the highest
weight
$\Lambda_{n+1}$ which in the
$E_{11}$ Dynkin diagram of figure 3 corresponds to node eleven [39].
This result only applies to terms that occur in the eleven
dimensional theory. ÊÊThe calculation of this paper, Êand that of
reference [39] also determines Êthe parameter $s$ Êof equation (3.3)
that
occurs in the automorphic form; for the ÊIIA and IIB theories we find
that
$s={l_T-2\over 4}$, ÊÊwhile for M theory we find that ÊÊ$s={l_T-2\over
6}$ where $l_T$ is the number of space-time derivatives in the term in
the effective action being considered.
\par
We will now consider if Êthe results just mentioned actually agree with
the known results in type II string theory. For
low numbers of space-time derivatives there are precise proposals for
the
automorphic forms that occur Êand their properties have been
checked Êagainst the known features of the perturbation
expansions of the type II strings [34, 35, 36,37]. One finds for the
$R^4$ term in $d \le 7 $ that the $E_{n+1}$ Êautomorphic form is
built Êfrom the representation with highest weight
$\Lambda_n$ and has
$s={3\over 2}$. This is completely consistent with the results found
from
the IIA and IIB viewpoints. For the $\partial^4 R^{4}$, or
equivalently
$R^6$, term in $d \le 7 $ the $E_{n+1}$ automorphic form is also built
from the Êrepresentation with highest weight $\Lambda_n$ and has
$s={5\over 2}$. However, Êin Ê$d=7$ dimensions the coefficient of this
term is in fact a sum of two $E_4=SL(5)$ automorphic terms
[35,36,37], in addition to an automorphic form constructed from the $
{\bar 5}$ of $SL(5)$, with $s={5 \over 2}$ one finds an automorphic
form built
from the $\bar{10}$ of $SL(5)$ with $s={5 \over 2}$. ÊSimilarly, in
$d=6$ dimensions the coefficient of the $\partial^4 R^{4}$ term is
the sum of an automorphic form constructed from the $10$ of $SO(5,5)
$, with $s={5 \over 2}$ and another automorphic form built from the
16-dimensional representation of $SO(5,5)$ with $s=3$. As these
additional automorphic Êforms disappear in the limits being
considered, the known automorphic forms Êfor the
$\partial^4 R^{4}$ term are also consistent with the results Êfound
from the
dimensional reduction of the type IIA and type IIB theories.
\par
However, dimensional reduction of the higher derivative correction of
the
eleven dimensional ÊÊtheory [39] Êsuggests that the automorphic forms
are
constructed from the representation with highest weight $\Lambda_{n+1}$.
At first sight this is inconsistent with the
automorphic forms that are known to be present. However, in seven
dimensions, i.e. for $SL(5)$, for the
$R^4$ term Êthis would imply in particular that the automorphic form
constructed from the $ \bar 5$ of $SL(5)$ with
$s={3\over 2}$ is proportional to the automorphic form constructed
from the $5$ of $SL(5)$ with
$s=1$. In fact this relation follows from the observation that an
automorphic form constructed from a given representation and another
automorphic form constructed from the corresponding Cartan involution
twisted representation are related by two suitable values of $s$
[34]. The
same holds for the automorphic forms associated with the $R^4$ terms in
lower dimensions as one knows [37] that the automorphic form
constructed from the representation of $E_{n+1}$ with highest weight $
\Lambda_{n}$ and
$s={3\over 2}$, i.e
$\Phi^{E_{n+1}}_{\Lambda_n ; {3\over 2}}$ is proportional to the
automorphic form constructed from the representation of $E_{n+1}$
with highest weight $\Lambda_{n+1}$ and $s=1$,
i.e
$\Phi^{E_{n+1}}_{\Lambda_{n+1}; 1}$, that is Ê[37]
$$
\Phi^{E_{n+1}}_{\Lambda_n; {3\over 2}}\propto
\Phi^{E_{n+1}}_{\Lambda_{n+1}; {1}}.
\eqno(6.1)$$
Some examples of relationships of this type were also found in
reference [31].
Consequently, Êthe known automorphic forms that occur for the $R^4$ term
are also in agreement with the prediction from the M theory viewpoint.
However, one can not apply the M theory results ÊÊto the
$R^6$ term as this term does not occur in the higher derivative
effective
action in eleven dimensions and so is not included in
the analysis from the M theory viewpoint given in [39]. Indeed,
the only terms that occur in eleven dimensions that involve, for
example, Êthe Riemann curvature are of the form
$R^{3n+1}$, for $n$ a positive integer.
\par
Given the above discussion, it is tempting to suppose the following

\item {[A]} The automorphic forms that occur as coefficients of the
higher
derivative terms in the string theory effective action must contain an
automorphic form constructed from the $\Lambda_n$ representation
of $E_{n+1}$.

\item {[B]} The automorphic forms Êthat occur in string theory ÊÊand
built from the
$\Lambda_{n}$ representation Êof $E_{n+1}$ are the same as the
automorphic
forms built
from the
$\Lambda_{n+1}$ representation of $E_{n+1}$ Êup to a numerical factor.

The first statement is phrased so as to allow for the possibility that
the coefficient is a sum of automorphic forms one or more of which may
disappear in the limit. ÊThe second Êstatement only
applies to automorphic forms of higher derivative terms that occur in
eleven dimensions.
\par
The automorphic
forms that are used in the recent work of [35-37] are those that
appear in
the work of Langlands, and they are eigenfunctions of the Laplacian and
the higher Casimir operators of
$E_{n+1}$. However, those that are constructed
in equation (3.3) are not in general
eigenfunctions of these operators. However,
one can impose constraints on the representations used
to construct the automorphic forms ÊÊand they then do become
eigenfunctions of the Laplacian and higher Casimir operators. This has
been worked out explicitly for the case of Êsix dimensions, i.e. for
$SO(5,5)$ with the ten dimensional vector representation where the
constraint
is that the length squared of this vector vanishes. Indeed only if this
constraint is implemented is the perturbation series in agreement with
that found in string theory; this part of the automorphic form
has been checked in detail to agree with the $SO(5,5)$ Langlands
automorphic
form for this representation [34]. It remains, however, to carry out
the
analogue of this construction for the higher rank groups and
representations. It is interesting to note that at least the
constant part of the Langlands automorphic forms can be written as a sum
of the Weyl group and this, being a rotation, Êpreserves the lengths of
vectors and those vectors that do occur must Êbelong to Êa single orbit.
As such, Êit is likely that the Langlands automorphic forms will involve
constraints on the representations used and will agree with the
automorphic forms of equation (3.3) once one imposes the appropriate
constraints.
\par
As we have mentioned, the detailed studies of the automorphic
forms in the low energy effective action of type II string theory
have only
concerned terms which have low numbers of space-time derivatives.
However, Êit is known that the automorphic forms that
occur Êas coefficients of the higher derivative terms Êin ten
dimensions that have more
than
twelve space-time derivatives, Êare not eigenvalues of the Laplacian and
so they can not be the ÊEisenstein automorphic forms found say in the
Langlands papers [37]. ÊÊAs a result the automorphic forms that
occur
for these higher derivative terms are Êessentially unknown. This
paper and
reference [39] puts some constraints on these objects. ÊWe have tacitly
assumed that all of the automorphic forms that appear as the
coefficient functions of the higher derivative terms are constructed
from a
representation of $E_{n+1}$. Although the form of equation (3.3) may not
be correct in general, even with constraints, the automorphic forms
will still have a dominant behaviour of the form Ê$e^{-s w\phi}
$ in
the limit Êstudied in this paper, Êso they will contain Êa parameter
$s$.
\par
We will now comment on the significance of the representations that
occur in the automorphic forms. ÊThe brane charges of
type II string theory in $d$-dimensions belong to representations of
$E_{n+1}$. ÊÊIn fact, there is very substantial evidence to believe
that all brane charges belong to the
$l_1$ representation of
$E_{11}$. Carrying out the decomposition of the $l_1$ representation we
find the brane charges ÊÊÊin
$d$ dimensions; they Êare listed in table one [49-51]. ÊThe first
entries of
the table agree with that found earlier using U duality transformations
[52]. Examining the table we find that the string charges, i.e Ê$Z^a$,
are
in the
$\Lambda_n$ representation, the membrane charges, i.e Ê$Z^{ab}$, are in
the
$\Lambda_{n+1}$ representation and the point particle Êcharges, i.e $Z$,
are in the
$\Lambda_1$ representation. ÊThus the above propositions can be
expressed Êas

\item {[A]}The automorphic forms that occur as coefficients of the
higher
derivative terms in the string theory effective action are constructed
from the string charge representation. We may very generically
write these automorphic forms as $\Phi_{string}$.

and that

\item {[B]} The automorphic forms Êthat occur in string theory
built Êfrom the
string charge representation are the same as the automorphic forms built
from the
membrane charge representation, ÊÊÊup to a numerical factor. We may
very generically write this as $\Phi_{string}=\Phi_{membrane}$.

As before the latter proposition only applies to the terms that have an
eleven-dimensional origin. It is of course very natural that the string
and membrane charge representations found in the automorphic forms arise
from the dimensional reduction of the ten dimensional IIA and IIB
string theories and the eleven dimensional theory respectively.
\par
It was also observed in reference [37] that the automorphic form for the
$R^4$ term are related to those built from the $\Lambda_1$
representation
as follows
$$
\Phi^{E_{n+1}}_{\Lambda_n; {3\over 2}}\propto
\Phi^{E_{n+1}}_{\Lambda_{1}; {n-2\over 2}}
\eqno(6.2)$$
for $n=4,5,6,7$ while for the $R^6$ term
$$
\Phi^{E_{n+1}}_{\Lambda_n; {5\over 2}}\propto
\Phi^{E_{n+1}}_{\Lambda_{1}; {n+2\over 2}}
\eqno(6.3)$$
for $n=4,5,6,7$.

Since the charges for the point particle belong to the $\Lambda_1$
representation Êwe are Êalso tempted to propose Êthat

\item {[C]}The automorphic forms Êthat occur in string theory ÊÊare
built Êfrom the
string charge representation are the same as the automorphic forms built
from the Êpoint charge representation ÊÊup to a numerical factor. We may
generically write this as
$\Phi_{string}=\Phi_{point}$.
\par
For the case of $d=7$ with the group $SL(5)$ this would require that the
automorphic forms constructed from the $\bar 5$ and $10$ representations
were the same for appropriate representations. In fact the automorphic
forms constructed by Langlands for the Êtwo representations $\Lambda
$ and
$\Lambda' $ are proportional Êif the vectors $\lambda = 2s\lambda -\rho$
and
$\lambda' = 2s'\lambda' -\rho$ Êare related by a Weyl reflection. ÊÊThe
Weyl vector $\rho$ can be written as
$\rho=\sum_a \Lambda_a$ where $\Lambda_a$ are the fundamental weights.
For our case Êwe should take $\Lambda=\Lambda_3$ and $\Lambda'=
\Lambda_1$. ÊSince Weyl reflections are rotations they preserve the
length squared and one finds that $\lambda^2= \lambda^{\prime 2}$ for
$s={3\over 2}$ if
$s'=2$ or $s'={1\over 2}$ and Êfor Ê$s={5\over 2}$ if
$s'={5\over 2}$. Indeed one can show that for $s={3\over 2}$
and Ê$s'={1\over 2}$ and Êalso for Ê$s={5\over 2}=s'$ Êthere is a Weyl
reflection of the Êrequired kind and so the relations of equations (6.2)
and (6.3) do extend to the case of $n=3$ are required. This is Êmost
easily found by writing the vectors ÊÊ$\lambda$ and $\lambda'$ in terms
of the orthonormal basis $e_a, a=1,2,3,4,5$ in terms of which the simple
roots take the form $\alpha_a=e_a-e_{a+1}$. As Weyl reflections permute
the $e_a$ basis it is straightforward to see if the two vectors are
related by a Weyl reflection.
\par
The presence of the highest weights $\Lambda_n$ and $\Lambda_{n+1}$
in the automorphic forms was deduced Êfrom dimensional reduction from
the
IIA (or IIB) theories and M theory respectively, but as we noted above
these representations correspond to nodes that are among those
deleted to find
these theories as non-linear realisations of $E_{11}$. As such one may
suspect that dimensionally reducing the theory from $d+1$ to $d$
dimensions will lead to the constraint that the automorphic form will
contain the
$\Lambda_1$ representation.

\medskip
{\bf 7. $E_{11}$ automorphic forms and Êhigher derivative corrections}
\medskip
The $E_{11}$ conjecture [43] involves a particular real form of $E_
{11}$ and
has so far been applied to the low energy Êeffective actions of string
theory, that is, the supergravity theories. As such it Êinvolves taking
this form of $E_{11}$ ÊÊover the real numbers. ÊHowever, we know
that $E_{11}$ Êrotates the brane charges, and as these are quantised
[12,13], for the full quantum string theory, we
should only consider a version of
$E_{11}$ which is over Êa discrete field rather than the real
numbers. In
particular, Êit should preserve the brane charge lattice
which
belongs to the $l_1$ representation [56,48,49,50]. We note that, since
$E_{11}$ includes the Lorentz group even this subgroup should be
taken over a discrete field. As such, Êit is not clear, at least at
first
sight, how ÊÊthe
$E_{11}$ conjecture can apply to the higher derivative terms and
how the theory of non-linear realisations Êcan be
applicable?
\par
The way the scalar fields occur in the low energy effective actions,
i.e.
supergravity theories, is Êcontrolled by the fact that they belong to a
non-linear realisation. The Ênon-linear realisation
for a group $G$, which does not involve space-time generators, with Êa
local subgroup $H$ is Êconstructed Êfrom a group element $g(\xi )$ which
is a function of Êspace-time and is subject to the
transformations
$g(\xi )\to g_0 g(\xi )$ and $g(\xi )\to gh(\xi )$ Êwhere $h$ belongs to
the local subalgebra $H$ and is also a Êfunction of
space-time and Ê$g_0$ is just an element of the Êgroup $G$ which is
independent of space-time. The
$\xi$ parameterise the group element and in the context of the
supergravity theories these are the scalars of the theory which
are themselves functions of space-time. ÊThis statement is true for all
scalar fields that belong to the supergravity multiplet; ÊÊfor the type
II theories in $d=10-n$ dimensions
$G=E_{n+1}$.
\par
In Êsupergravity theories the space-time derivatives of the
scalars are contained in the Cartan forms, while Êthe space-time
derivatives of the other fields occur together with the scalar fields in
just such a Êway that it converts the linear representations to which
they
belong into non-linear representations using the group element $g(\xi)$.
This construction has already been used in this paper and is
described in the appendix A.
\par
As we have mentioned the higher derivative terms of the string effective
action in
$d=11-n-1$ dimensions are conjectured to be invariant under a discrete
$E_{n+1}$ symmetry. In the effective action the Êspace-time derivatives
of the scalars and all the other fields occur in
precisely the same way as the supergravity theories; Êthe space-time
derivatives of the scalars appear as part of the Cartan forms and the
derivatives of the other fields arise in a non-linear representation of
$E_{n+1}$. ÊHowever, ÊÊeach term in the higher derivative effective
action
can have Êa coefficient that is an Êautomorphic form rather than a
constant as is the case of the supergravity theories.
The Êautomorphic forms of the
type, briefly described around Êequation (3.3), and used in references
[34] Êand also the automorphic forms of Langlands used in [35-37] Êare
constructed from a given representation of
$G$ using precisely the same group element
$g_{E_{n+1}}(\xi )$ except for the fact that the
$g_0$ transformation is now Êover the corresponding discrete group.
Indeed
the automorphic form is constructed from the non-linear
representation, that is,
$|\varphi>= I(g^{-1}_{E_{n+1}})|\psi>$ where the $| \psi >$ carry the
linear representation.
Hence Êalthough the higher derivative string effective action is only
invariant under a discrete
$E_{n+1}$ Êsymmetry Êit is
constructed using much of the same machinery as the $E_{n+1}$ Ênon-
linear
realisation that arises in ÊÊthe supergravity theories.
\par
Let us Ênow Êexamine if the building blocks
used to construct the higher derivative effective action appear Êin the
non-linear realisation of
$E_{11}$ appropriate to $d$ dimensions. The $E_{11}$ group element
has the
generic form [43,44,45]
$$
g_E= e^{h_a{}^b K^a{}_b} e^{A\cdot R}\ldots Êg_{E+1}(\xi)
\eqno(7.1)$$
where $g_{E}(\xi)$ contains the scalar fields and it is the group
element used in the non-linear realisation of $E_{n+1}$ as just
mentioned
in the previous paragraph. ÊThe Cartan forms of the $E_{11}$ group
element
of equation (6.4) Êcontain the
$E_{n+1}$ Cartan form of the scalars and the derivatives of the
other fields Êas Ênon-linear representations of $E_{n+1}$.
As such the $E_{11}$ non-linear realisation contains all the building
blocks of the higher derivative effective action, Êincluding the group
element $g_{E+1}$ Êwhich was used to construct the
automorphic forms; Êthe one exception is the
representation used to construct the automorphic forms.
\par
Space-time is introduced Êinto Êthe $E_{11}$ theory by considering Êthe
fundamental representation of $E_{11}$ associated with node one, denoted
$l_1$ [56]. In particular one takes the non-linear realisation of the
semi-direct product of $E_{11}$ and generators that belong to the $l_1$
representation, denoted $E_{11}\otimes_s l_1$. The corresponding group
element Êis given by
$g=g_l g_E$ Êwhere $g_E$ is the group element of $E_{11}$, given Êin
equation (7.1), and
$g_l$, in $d$ Êdimensions, Êis of the form [56]
$$
g_l= e^{x^a P_a} e^{z\cdot Z} e^{z_{a }\cdot Z^{a}} e^{z_{ab }\cdot
Z^{ab}}
\ldots
\eqno(7.2)$$
where $P_a, Z, Z^{a},\ldots $ are the generators that belong to the
$l_1$
representation decomposed into representations of $GL(d)\otimes
E_{n+1}$.
In particular
$P_a$ are the space-time translations in
$d$ dimensions, $Z$ are the scalar, that is point particle, Êcharges,
$Z^{a}$ are the string charges, $Z^{ab}$ the membrane charges $\ldots $
etc. These charges belong to the
$\Lambda_1$, $\Lambda_n$, $\Lambda_{n+1}, \ldots $
representations of
$E_{n+1}$ [49-51], see table one.
The $x^a, z, z_a, z_{ab},\ldots $ are the coordinates of the generalised
space-time.
As Êthe $l_1$ representation ÊÊcontains all the brane charges
there is a correspondence between the coordinates of Êthe generalised
space-time and Êthe brane charges. ÊThe
non-linear realisation $E_{11}\otimes_s l_1$ ÊÊlargely Êspecifies Êthe
generalised geometry corresponding to this generalised space-time.
\par
It is intriguing that the point particle,
string and membrane representations of $E_{n+1}$ Êcontained in Êthe
$l_1$
representations are just the ones that show up in the automorphic forms
that occur in string theory. Taking this Êtogether with our previous
comments Êwe find that all the ingredients of higher derivative
corrections can be found in the non-linear realisation of
$E_{11}\otimes l_1$. ÊThis
is at least consistent with the possibility of an $E_{11}\otimes
l_1$ formulation of the higher derivative effective action. To construct
$E_{11}$ automorphic forms using the $l_1$ representation is
straightforward, at least at the naive level,
as the construction of reference [34] can be applied
straightforwardly. Furthermore as we have explained above one can write
the space-time derivatives of the fields as part of the $E_{11}$ Cartan
forms. When considering the non-linear realisation of $E_{11}\otimes_s
l_1$ one would expect the Êparts involving the usual space-time
derivatives, that is the derivatives of the fields, Êto rotate into the
parts involving the other parts of the
$l_1$ representation, that is the parts in the automorphic form.
However, it is not immediately clear how to construct all of Êthe
higher
derivative Êstring effective action from the $E_{11}$ non-linear
realisation. While one might think that an $E_{11}$ invariant exists at
each order in space-time derivatives one could dream that there is an
enlargement Êof the symmetry algebra that combines all orders in
derivatives into a single automorphic form. ÊAnother point to bear in
mind
is that the
$E_{11}\otimes_s l_1$ non-linear realisation leads in the context of
supergravity theories, to equations of motion that contain only one
space-time derivative and it is Ênot apparent Êhow this could be
generalised to incorporate the higher derivative corrections.
\par
The remaining problem in applying $E_{11}$ to the low energy effective
action is how to reconcile the generalised space-time encoded in the
$E_{11}\otimes_s l_1$ non-linear realisation with the usual formulations
of supergravity Êthat involves just the conventional
coordinates $x^\mu$ of space-time. The work of reference [53] suggested
that even though Êthe full theory was
$E_{11}\otimes l_1$ invariant only part of the $l_1$ representation
occurred in the second quantised field theory. In particular although
the first quantised theory involved all of the $l_1$ representation, the
choice of representation of the commutators that takes one to get
from the
first to the second quantised theory required one to choose only part of
the $l_1$ representation. However, one can Êmake different choices of
which part of the $l_1$ representation one takes and these should be
equivalent and related by
$E_{11}$ transformations. The fact that Êautomorphic forms constructed
from different representations are the same and so lead to the same
theory Êcould be consistent with the above observation. Indeed one can
view it as a kind of uncertainty principle.
\par
We note another similarity between the higher derivative effective
action
and the $E_{11}\otimes_s l_1$ non-linear realisation.
Dimensionally reducing a field theory on an internal space and keeping
all the Kaluza-Klein modes just expresses the original theory on the
internal space Êin an alternative, but equivalent Êform. The higher
derivative corrections do keep a knowledge of the Kaluza-Klein modes as
they Êlead to some of the integer sums that occur
in the automorphic forms. Indeed, one can wonder if it is possible to
reconstruct
the higher derivative corrections Êin $d+1$ dimensions from those in $d$
dimensions. This is a feature the higher derivative corrections share
with the $E_{11}\otimes_s l_1$ non-linear realisation where the
different
theories, that is IIA , M theory, IIB and the theories in
$d\le 10 $ dimensions, just correspond to different decompositions of
$E_{11}\otimes_s l_1$; the count of fields and coordinates being the
same in the different theories.
\par
Finally, we close this section by noting the underlying significance of
determining Êthe automorphic forms. The type II supergravity theories
are
the complete low energy effective actions for the type II string. The
complete Êeffective action would be known if we knew Êthe automorphic
forms that occur as coefficients Êof the individual terms. However,
these
automorphic forms are not the ones traditionally studied Êin that they
are
non-holomorphic. For the automorphic forms that arise as coefficients of
terms with a Êlow numbers of space-time derivatives, Êtheir
non-holomorphic character is compensated for Êby the fact that they are
eigenvalues of the Laplacian Êand higher order Casimir Êoperators.
However, as we previously mentioned little is known about the
automorphic
forms that occur in general, but one should expect that they Êdo have a
defining characteristic. Such a knowledge would allow one to completely
specify all effects Êof ÊÊtype II string theory, ÊÊat least compactified
on a torus.
\vfil
\eject
$$\halign{\centerline{#} \cr
\vbox{\offinterlineskip
\halign{\strut \vrule \quad \hfil # \hfil\quad &\vrule Ê\quad \hfil #
\hfil\quad &\vrule \hfil # \hfil
&\vrule \hfil # \hfil Ê&\vrule \hfil # \hfil &\vrule \hfil # \hfil &
\vrule \hfil # \hfil &\vrule \hfil # \hfil &\vrule \hfil # \hfil &
\vrule \hfil # \hfil &\vrule#
\cr
\noalign{\hrule}
D&G&$Z$&$Z^{a}$&$Z^{a_1a_2}$&$Z^{a_1\ldots a_{3}}$&$Z^{a_1\ldots a_
{4}}$&$Z^{a_1\ldots a_{5}}$&$Z^{a_1\ldots a_6}$&$Z^{a_1\ldots a_7}$&\cr
\noalign{\hrule}
8&$SL(3)\otimes SL(2)$&$\bf (3,2)$&$\bf (\bar 3,1)$&$\bf (1,2)$&$\bf
(3,1)$&$\bf (\bar 3,2)$&$\bf (1,3)$&$\bf (3,2)$&$\bf (6,1)$&\cr
&&&&&&&$\bf (8,1)$&$\bf (6,2)$&$\bf (18,1)$&\cr Ê&&&&&&&$\bf (1,1)$&&$
\bf
(3,1)$&\cr Ê&&&&&&&&&$\bf (6,1)$&\cr
&&&&&&&&&$\bf (3,3)$&\cr
\noalign{\hrule}
7&$SL(5)$&$\bf 10$&$\bf\bar 5$&$\bf 5$&$\bf \bar {10}$&$\bf 24$&$\bf
40$&$\bf 70$&-&\cr Ê&&&&&&$\bf 1$&$\bf 15$&$\bf 50$&-&\cr
&&&&&&&$\bf 10$&$\bf 45$&-&\cr
&&&&&&&&$\bf 5$&-&\cr
\noalign{\hrule}
6&$SO(5,5)$&$\bf \bar {16}$&$\bf 10$&$\bf 16$&$\bf 45$&$\bf \bar
{144}$&$\bf 320$&-&-&\cr &&&&&$\bf 1$&$\bf 16$&$\bf 126$&-&-&\cr
&&&&&&&$\bf 120$&-&-&\cr
\noalign{\hrule}
5&$E_6$&$\bf\bar { 27}$&$\bf 27$&$\bf 78$&$\bf \bar {351}$&$\bf
1728$&-&-&-&\cr Ê&&&&$\bf 1$&$\bf \bar {27}$&$\bf 351$&-&-&-&\cr
&&&&&&$\bf 27$&-&-&-&\cr
\noalign{\hrule}
4&$E_7$&$\bf 56$&$\bf 133$&$\bf 912$&$\bf 8645$&-&-&-&-&\cr
&&&$\bf 1$&$\bf 56$&$\bf 1539$&-&-&-&-&\cr
&&&&&$\bf 133$&-&-&-&-&\cr
&&&&&$\bf 1$&-&-&-&-&\cr
\noalign{\hrule}
3&$E_8$&$\bf 248$&$\bf 3875$&$\bf 147250$&-&-&-&-&-&\cr
&&$\bf1$&$\bf248$&$\bf 30380$&-&-&-&-&-&\cr
&&&$\bf 1$&$\bf 3875$&-&-&-&-&-&\cr
&&&&$\bf 248$&-&-&-&-&-&\cr
&&&&$\bf 1$&-&-&-&-&-&\cr
\noalign{\hrule}
}}\cr
}$$

\par

\centerline {Table 1. The Brane ÊCharge representations of
the group, G, derived from the $l_1$ representation of $E_{11}$ [49-51]}

\medskip
{\bf Acknowledgment}
\medskip
The authors wishes to thank Neil Lambert Êfor discussions.
Peter West Êalso thanks the STFC for support from the Êgrant
awarded to theory group at King's.

\bigskip
{\bf {Appendix A: Non-linear Realisations}}
\bigskip

In this appendix we review the Êconstruction of non-linear
realisations in a form suitable to that used in this paper. We
consider a group $G$ with Lie algebra $Lie(G)$. $Lie(G)$ can be
split into the Cartan subalgebra with elements $\vec H$, positive
root generators $E_{\vec\alpha}$ and negative root generators
$E_{-\vec\alpha}$ with $\vec\alpha>0$. There exists a natural
involution, known as the Cartan involution, defined by
$$ \tau
:(\vec H,E_{\vec\alpha})\to -(\vec H, E_{-\vec\alpha})\ . \eqno
(A.1)
$$
To construct the Ênon-linear realisation we must specify a
subgroup $H$ (not to be confused with Êthe generators of the Cartan
subgroup which are denoted by $\vec H$). For us this is defined to
be Êthe subgroup left invariant under the Cartan involution, i.e.
$H=\{g\ \in \ G:\tau(g)=g \}$. In terms of the Lie algebra $Lie(H)$
it is all elements $A$ such that $A=\tau(A)$.

The non-linear realisation is constructed from group elements $g(x)
\in G$ that depend on spacetime that are
subject to the transformations
$$
g(x)\to g_0 g(x) h^{-1}(x)\ , \eqno (A.2)
$$
where $g_0\in G$ is constant and $h(x)\in H$ depends on spacetime. We
may write the group element in the form
$$
g(x) = e^{\sum_{\vec\alpha>0} \chi_{\vec\alpha} E_{\vec\alpha}}
e^{-{1\over \sqrt{2}}\vec\phi\cdot\vec H} e^{\sum_ {\vec\alpha>0}
u_{\vec\alpha} E_{-\vec\alpha}}\ ,\eqno(A.4)
$$
but using the local transformation we can bring it
to the form
$$
g(\xi) = e^{\sum_{\vec\alpha>0} \chi_{\vec\alpha} E_{\vec\alpha}}
e^{-{1\over \sqrt{2}}\vec\phi\cdot\vec H}\ .\eqno(A.5)
$$
Here we use Ê$\xi = (\vec\phi,\chi_{\vec\alpha})$ as a generic
symbol for all the scalar fields, which are functions of spacetime,
that parameterize the coset representative. Under a rigid $g_0\in G$
transformation $g(\xi) \to g_0 g(\xi)$ this form for the coset
representative is not preserved. However one can make a compensating
transformation $h(g_0,\xi)\in H$ that returns $g_0g(\xi) $ into the
form of equation (A.5);
$$
g_0 g(\xi)h^{-1}(g_0,\xi) = g(g_0\cdot \xi)\ .\eqno (A.6)
$$
This induces a non-linear action of the group $G$ on
the scalars; $\xi \to g_0\cdot \xi$.

We will also need Êa Êlinear representation of $G$. ÊLet
$\vec\mu^i$, $i=1,...,N$ be the weights ÊÊof the representation and
$|\vec\mu^i>$ be a corresponding Êstates. We choose $\vec\mu^1$ to
be Êthe highest weight and so the corresponding state satisfies
$E_{\vec\alpha}|\vec\mu^1>=0$ for all simple roots $\vec\alpha$.
The states in the rest of the representation are polynomials of
$F_{\vec \alpha}= E_{-\alpha}$ acting on the highest weight state.

We consider states of the form Ê$|\psi>=\sum_i\psi_i |\vec\mu^i>$.
Under the action $U(g_0)$ of the group $G$ we have
$$
|\psi>\to U(g_0)|\psi>=L( g_0^{-1}) \sum_i \psi_i|\vec\mu^i> \equiv
(U (g_0)\psi_i) |\vec\mu^i>= \sum_{i,j}
D_i{}^j (g_0^{-1})\psi_j |\vec\mu^i> \ ,\eqno (A.7)
$$
where $L(g_0) $ is the expression of the group element $g_0$ in terms
of the Lie algebra elements which now act
on the states of the representation in the usual way. We note that
the action of the group on the components
$\psi_i$ is given by $ \psi_i\to U(g_0)\psi_i= \sum_j D_i{}^j (g_0^
{-1})\psi_j $ which is the result expected
for a passive action. The advantage of using the states to discuss
the representation is that Êwe can use the
action of the Lie algebra elements $L(g_0) $ on the states to compute
the matrix $ D_i{}^j$ of the
representation and Êdeduce properties of the representation in general.

Given any linear realisation, we
can construct a non-linear realisation by
$$
|\varphi(\xi)> = \sum \varphi_i |\vec\mu^i> =L(g^{-1}(\xi))|\psi> =
e^{ {1\over \sqrt{2}}\vec\phi\cdot\vec
H}e^{-\sum_{\vec\alpha>0}\chi_{\vec \alpha} E_{\vec\alpha}} |\psi>
\ ,\eqno (A.8)
$$
where $g(\xi)$ is the group element of the Ênon-linear realisation
in equation (A.5). Under a group transformation $U(g_0)$ it
transforms as
$$\eqalign{
U(g_0)|\varphi(\xi)> &= ÊL(g^{-1}(\xi))U(g_0)|\psi> =L(g^{-1}(\xi)) L
(g_0^{-1} ) |\psi> \cr &= L( (g_0
g^{-1}(\xi))|\psi>\cr &= L(h^{-1})|\varphi(g_0\cdot\xi)> \ ,}\eqno (A.9)
$$
using equation (A.2). In terms of the component fields we find that $
\varphi_i (\xi)= \sum_j
D_i{}^j(g^{-1}(\xi))\psi_j$ and $U(g_0)\varphi_i (\xi)= \sum_j D_i{}^j
((h)^{-1})\varphi_j (g_0\cdot \xi)$.

\bigskip
{\bf {Appendix B: The decomposition of the simple roots and weights of
$E_{n+1}$ }}
\bigskip

In this paper we carry out the decomposition of certain
representations of
$E_{n+1}$ into those of $SL(n)\otimes GL(1) \otimes GL(1)$. ÊThe $E_{n
+1}$
algebra that appears in the dimensional reduction of the IIA theory
on an
$n-1$ torus to $d$ dimensions appears, from an $E_{11}$ perspective by
deleting node $d$ in the $E_{11}$ dynkin diagram.
$$
\matrix {
& & & & & & & & & 11 & & Ê&&\cr
& & & & & & & & Ê& \bullet & & && \cr
& & & & & & & & & | & & &&\cr
\bullet & Ê- & Ê\ldots & -
&\otimes&-&\ldots -&\bullet&-&\bullet&-&\bullet&-&\bullet
\cr 1& & & & d & & &7 Ê& Ê& 8 & Ê& 9&&10 }
$$
\medskip
\centerline {Figure 4. The $E_{11}$ Dynkin diagram appropriate to the}
\centerline {$d$-dimensional maximal supergravity theory}
\medskip
The Dynkin diagram of the remaining $E_{n+1}$ internal subalgebra
resulting from deleting node $d$ in figure 4, where nodes $(d+1)$,...,
$11$, become $1$,...,$(n+1)$ is then given in figure 5.
$$
\matrix {
& & & & & n+1 & & Ê&&\cr
& & & & Ê& \bullet & & && \cr
& & & & & | & & &&\cr
\bullet &-&\ldots -&\bullet&-&\bullet&-&\bullet&-&\bullet \cr
1 Ê& & &n-3 Ê& Ê& n-2 & Ê& n-1& &n }
$$
\medskip
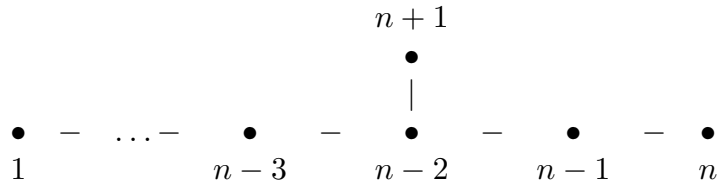
\centerline {Figure 5. The $E_{n+1}$ Dynkin diagram of the internal
subalgebra}
\centerline {arising from deleting node $d$ in the $E_{11}$ Dynkin
diagram}
\medskip

However, we can also delete nodes
$n$ and $n+1$ of this Dynkin diagram. The
deletion of nodes $n$ and $n+1$ leads to
the algebra $SL(n)\otimes GL(1)\otimes GL(1)$. In this appendix we will
find how the roots and weights of $E_{n+1}$ in terms of those of
$SL(n)\otimes GL(1)\otimes GL(1)$.
\par
Let us carry out the decomposition by Êfirst
deleting Ênode
$n$ to find the roots and fundamental weights of $D_{n}$ and then
delete node
$n+1$ to find the algebra $SL(n)$. Using the methods given
in reference [54], the simple roots of
$E_{n+1}$ can be expressed as
$$
\vec \alpha_{i} = \left(0, \underline {\tilde \alpha}_{i} \right),
\quad i
= 1,...,n-1, n+1 \quad
\vec \alpha_{n}=\left(x,- Ê\tilde{\underline \lambda}_{n-1} \right)
\eqno(B.1)$$
Here
$\underline {\tilde\alpha} _{i}, i =1,...,n$ are the roots of $D_
{n}$
and
$\underline {\tilde \lambda}_i$ its fundamental weights which are given
by
$$
\vec \Lambda_{i} = \left({\underline {\tilde \lambda}_i\cdot \underline
{\tilde \lambda} _{n-1}\over x},
\underline
{\tilde \lambda}_i\right),
\quad
i = 1,...,n-1, n+1 \quad
\vec \Lambda_{n}=\left({1\over x}, \underline 0 \right)
\quad\
\eqno(B.2)$$
The variable
$x$ is fixed by demanding that $\alpha_{n}^2=2=x^2+\underline
{\tilde \lambda} _{n-1}^2$.
\par
We now delete node $n$ to find the $A_{n-1}$ algebra. The roots of
$E_{n+1}$ are found from the above roots by substituting the
corresponding
decomposition of the $D_{n}$ roots and weights into those of $A_
{n-1}$.
The roots of $D_{n}$ in terms of those of $A_{n-1}$ are given by
$\underline {\tilde \alpha}_i = \left(0, \underline \alpha_{i}\right),\
i=1,...,n -1$ and
$\underline {\tilde \alpha}_{n} = \left(y,- \underline
\lambda_{n-2}\right)$ while the fundamental weights are given by
$\underline {\tilde \lambda}_i = \left({\lambda _{n-2}\cdot
\lambda_{i}\over y},
\underline
\lambda_{i}\right)\ i=1,...,n -1$ and
$\underline {\tilde \lambda}_{n+1} = \left({1\over y}, \underline
0\right)$. Requiring
${\tilde \alpha}_{n+1}^2=2$ gives $y^2={4\over n}$
We then find that the roots of $E_{n+1}$ are given by
$$
\vec \alpha_{i} = \left(0, 0,\underline \alpha_{i} \right), \quad
i =1,...,n-1,
$$
$$
\vec \alpha_{n}=\left(x , -{\lambda _{n-2}\cdot \lambda_{n-1}\over
y}, -\underline ÊÊÊ\lambda_{n-1}
\right),
$$
$$
\vec \alpha_{n+1}=\left(0,y, -\underline \lambda_{n-2} \right)
\eqno(B.3)$$
\par
The fundamental weights of $E_{n+1}$ are found in the same way to be
$$
\vec{\Lambda}_{i} = \left( {c_i\over x},{\lambda _{n-2}\cdot
\lambda_{i}\over y},
\underline \lambda_{i} \right), Ê\quad i =1,...,n-1,
$$
$$
\vec{\Lambda}_{n} = \left( {1\over x},0 , \underline 0 Ê\right),
$$
$$
\vec{\Lambda}_{n+1} = \left( {n-2\over 4 x},{1\over y} , \underline 0
\right).
\eqno(B.4)$$
where $ c_i={i\over 2}, \ i=1,\ldots ,n-2$ and Ê$ c_{n-1}={n \over 4}$.
As Ê$\underline {\tilde
\lambda}_{n-1}^2={n\over 4}$ we find that $x^2={8-n\over 4}$.

\medskip
\noindent
{\bf References}
\medskip

\item{[1]}
ÊÊI.~C.~G.~Campbell and P.~C.~West,
ÊÊ{\it N=2 D=10 Nonchiral Supergravity and Its Spontaneous
Compactification},
ÊÊNucl.\ Phys. ÊB {\bf 243} (1984) 112.
ÊÊ

\item{[2]}
ÊÊF.~Giani and M.~Pernici,
ÊÊ{\it N=2 Supergravity in Ten-Dimensions},
ÊÊPhys.\ Rev. ÊD {\bf 30} (1984) 325.
ÊÊ

\item{[3]}
ÊÊM.~Huq and M.~A.~Namazie,
ÊÊ{\it Kaluza-Klein Supergravity in Ten-Dimensions},
ÊÊClass.\ Quant.\ Grav.\ {\bf 2}, 293 (1985)
ÊÊ[Erratum-ibid.\ Ê{\bf 2}, 597 (1985)].
ÊÊ
ÊÊ\medskip
\item{[4]}
ÊÊJ.~H.~Schwarz and P.~C.~West,
ÊÊ{\it Symmetries and Transformations of Chiral N=2 D=10 Supergravity},
ÊÊPhys.\ Lett. ÊB {\bf 126}, 301 (1983).
ÊÊ

\item{[5]}
ÊÊP.~S.~Howe and P.~C.~West,
ÊÊ{\it The Complete N=2, D=10 Supergravity},
ÊÊNucl.\ Phys. ÊB {\bf 238}, 181 (1984).
ÊÊ

\item{[6]}
ÊÊJ.~H.~Schwarz,
ÊÊ{\it Covariant Field Equations of Chiral N=2 D=10 Supergravity},
ÊÊNucl.\ Phys. ÊB {\bf 226}, 269 (1983).
ÊÊ

\item{[7]}
ÊÊE.~Cremmer, B.~Julia and J.~Scherk,
ÊÊ{\it Supergravity Theory in Eleven-Dimensions},
ÊÊPhys.\ Lett. B {\bf 76}, 409 (1978).
ÊÊ

\item{[8]}
ÊÊE.~Cremmer and B.~Julia,
ÊÊ{\it The N=8 Supergravity Theory. 1. The Lagrangian},
ÊÊPhys.\ Lett. ÊB {\bf 80}, 48 (1978).
ÊÊ

\item{[9]}
ÊÊN.~Marcus and J.~H.~Schwarz,
ÊÊ{\it Three-Dimensional Supergravity Theories},
ÊÊNucl.\ Phys. ÊB {\bf 228}, 145 (1983).
ÊÊ

\item{[10]}
ÊÊB.~Julia and H.~Nicolai,
ÊÊ{\it Conformal internal symmetry of 2-d sigma models coupled to
gravity and a
ÊÊdilaton},
ÊÊNucl.\ Phys. ÊB {\bf 482}, 431 (1996)
ÊÊ[arXiv:hep-th/9608082].
ÊÊ

\item{[11]}
ÊÊÊÊÊÊÊÊB.~Julia, in {\it Vertex Operators and Mathematical
Physics}, Publications of the Mathematical
ÊÊÊÊÊÊÊÊSciences Research Institute no3. Springer Verlag (1984); in
{\it Superspace and
ÊÊÊÊÊÊÊÊSupergravity}, ed. S. W. Hawking and M. Rocek, Cambridge
University Press (1981)

\item{[12]}
ÊÊC.~Teitelboim,
ÊÊ{\it Monopoles of Higher Rank},
ÊÊPhys.\ Lett. ÊB {\bf 167}, 69 (1986).
ÊÊ

\item{[13]}
ÊÊR.~I.~Nepomechie,
ÊÊ{\it Magnetic Monopoles from Antisymmetric Tensor Gauge Fields},
ÊÊPhys.\ Rev. ÊD {\bf 31}, 1921 (1985).
ÊÊ

\item{[14]}
ÊÊA.~Font, L.~E.~Ibanez, D.~Lust and F.~Quevedo,
ÊÊ{\it Strong - weak coupling duality and nonperturbative effects in
string
ÊÊtheory},
ÊÊPhys.\ Lett. ÊB {\bf 249}, 35 (1990).
ÊÊ
S.J. Rey, {\it The Confining Phase Of Superstrings And Axionic
Strings}, Phys. Rev. {\bf D43} (1991) 526.

\item{[15]}
ÊÊA.~Sen,
ÊÊ{\it Electric magnetic duality in string theory},
ÊÊNucl.\ Phys. ÊB {\bf 404}, 109 (1993)
ÊÊ[arXiv:hep-th/9207053].
ÊÊ
A. Sen, {\it Quantization of dyon charge and electric magnetic
duality in
string theory}, Phys. Lett. {\bf 303B} (1993) 22; {\it Strong - weak
coupling duality in four-dimensional string theory}, Int. J. Mod. Phys.
{\bf A9} (1994) 3707. J. Schwarz and A. Sen, {\it Duality symmetric
actions}, Nucl. Phys. {\bf B411} (1994) 35.

\item{[16]}
ÊÊC.~M.~Hull and P.~K.~Townsend,
ÊÊ{\it Unity of superstring dualities}
ÊÊNucl.\ Phys. ÊB {\bf 438}, 109 (1995)
ÊÊ[arXiv:hep-th/9410167].
ÊÊ

\item{[17]}
ÊÊM.~B.~Green and M.~Gutperle,
ÊÊ{\it Effects of D instantons},
ÊÊNucl.\ Phys. ÊB {\bf 498}, 195 (1997)
ÊÊ[arXiv:hep-th/9701093].
ÊÊ

ÊÊ
\item{[18]}
ÊÊM.~B.~Green, M.~Gutperle and P.~Vanhove,
ÊÊ{\it One loop in eleven-dimensions},
ÊÊPhys.\ Lett. ÊB {\bf 409}, 177 (1997)
ÊÊ[arXiv:hep-th/9706175].
ÊÊ

ÊÊ
\item{[19]}
ÊÊM.~B.~Green and S.~Sethi,
ÊÊ{\it Supersymmetry constraints on type IIB supergravity},
ÊÊPhys.\ Rev. ÊD {\bf 59}, 046006 (1999)
ÊÊ[arXiv:hep-th/9808061].
ÊÊ

ÊÊ
\item{[20]}
ÊÊM.~B.~Green, H.~h.~Kwon and P.~Vanhove,
ÊÊ{\it Two loops in eleven-dimensions},
ÊÊPhys.\ Rev. ÊD {\bf 61}, 104010 (2000)
ÊÊ[arXiv:hep-th/9910055].
ÊÊ

ÊÊ
\item{[21]}
ÊÊM.~B.~Green and P.~Vanhove,
ÊÊ{\it Duality and higher derivative terms in M theory},
ÊÊJHEP {\bf 0601}, 093 (2006)
ÊÊ[arXiv:hep-th/0510027].
ÊÊ

ÊÊ
\item{[22]}
ÊÊM.~B.~Green, J.~G.~Russo and P.~Vanhove,
ÊÊ{\it Modular properties of two-loop maximal supergravity and
connections with
ÊÊstring theory},
ÊÊJHEP {\bf 0807}, 126 (2008)
ÊÊ[arXiv:0807.0389 [hep-th]].
ÊÊ

\item{[23]}
ÊÊJ.~G.~Russo,
ÊÊ{\it Construction of SL(2,Z) invariant amplitudes in type IIB
superstring
ÊÊtheory},
ÊÊNucl.\ Phys. ÊB {\bf 535}, 116 (1998)
ÊÊ[arXiv:hep-th/9802090].
ÊÊ

\item{[24]}
ÊÊA.~Basu,
ÊÊ{\it The D**10 R**4 term in type IIB string theory},
ÊÊPhys.\ Lett. ÊB {\bf 648}, 378 (2007)
ÊÊ[arXiv:hep-th/0610335].
ÊÊ

\item{[25]}
ÊÊN.~Berkovits and C.~Vafa,
ÊÊ{\it Type IIB R**4 H**(4g-4) conjectures},
ÊÊNucl.\ Phys. ÊB {\bf 533}, 181 (1998)
ÊÊ[arXiv:hep-th/9803145].
ÊÊ

\item{[26]}
ÊÊJ.~G.~Russo and A.~A.~Tseytlin,
ÊÊ{\it One loop four graviton amplitude in eleven-dimensional
supergravity},
ÊÊNucl.\ Phys. ÊB {\bf 508}, 245 (1997)
ÊÊ[arXiv:hep-th/9707134].
ÊÊ

ÊÊ
\item{[27]}
ÊÊM.~B.~Green, J.~G.~Russo and P.~Vanhove,
ÊÊ{\it Modular properties of two-loop maximal supergravity and
connections with
ÊÊstring theory},
ÊÊJHEP {\bf 0807}, 126 (2008)
ÊÊ[arXiv:0807.0389 [hep-th]].
ÊÊ

ÊÊÊÊ
\item{[28]}
ÊÊM.~B.~Green, J.~G.~Russo and P.~Vanhove,
ÊÊ{\it Non-renormalisation conditions in type II string theory and
maximal
ÊÊsupergravity},
ÊÊJHEP {\bf 0702}, 099 (2007) [arXiv:hep-th/0610299].
ÊÊ

\item{[29]}
ÊÊA.~Basu,
ÊÊ{\it The D**4 R**4 term in type IIB string theory on T**2 and U-
duality},
ÊÊPhys.\ Rev. ÊD {\bf 77}, 106003 (2008)
ÊÊ[arXiv:0708.2950 [hep-th]];
ÊÊA.~Basu, {\it Supersymmetry constraints on the $R^4$ multiplet in
type IIB on $T^2$},
ÊÊarXiv:1107.3353 [hep-th].
ÊÊ
ÊÊ

\item{[30]}
ÊÊE.~Kiritsis and B.~Pioline,
ÊÊ{\it On R**4 threshold corrections in IIb string theory and (p, q)
string
ÊÊinstantons},
ÊÊNucl.\ Phys. ÊB {\bf 508}, 509 (1997)
ÊÊ[arXiv:hep-th/9707018].
ÊÊ

ÊÊ
\item{[31]}
ÊÊN.~A.~Obers and B.~Pioline,
ÊÊ{\it Eisenstein series and string thresholds},
ÊÊCommun.\ Math.\ Phys. Ê{\bf 209}, 275 (2000)
ÊÊ[arXiv:hep-th/9903113]; N.~A.~Obers and B.~Pioline,
ÊÊ{\it Eisenstein series in string theory},
ÊÊClass.\ Quant.\ Grav. Ê{\bf 17}, 1215 (2000)
ÊÊ[arXiv:hep-th/9910115].
ÊÊ

\item{[32]}
ÊÊN.~Lambert and P.~C.~West,
ÊÊ{\it Enhanced Coset Symmetries and Higher Derivative Corrections},
ÊÊPhys.\ Rev. ÊD {\bf 74}, 065002 (2006)
ÊÊ[arXiv:hep-th/0603255].
ÊÊ

ÊÊ
\item{[33]}
ÊÊN.~Lambert and P.~C.~West,
ÊÊ{\it Duality Groups, Automorphic Forms and Higher Derivative
Corrections},
ÊÊPhys.\ Rev. ÊD {\bf 75}, 066002 (2007)
ÊÊ[arXiv:hep-th/0611318].
ÊÊ

ÊÊ
\item{[34]}
ÊÊN.~Lambert and P.~West,
ÊÊ{\it Perturbation Theory From Automorphic Forms},
ÊÊJHEP {\bf 1005}, 098 (2010)
ÊÊ[arXiv:1001.3284 [hep-th]].
ÊÊ

\item{[35]}
ÊÊM.~B.~Green, J.~G.~Russo and P.~Vanhove,
ÊÊ{\it Automorphic properties of low energy string amplitudes in
various
ÊÊdimensions},
ÊÊPhys.\ Rev. ÊD {\bf 81}, 086008 (2010)
ÊÊ[arXiv:1001.2535 [hep-th]].
ÊÊ

\item{[36]}
ÊÊM.~B.~Green, J.~G.~Russo and P.~Vanhove,
ÊÊ{\it String theory dualities and supergravity divergences},
ÊÊJHEP {\bf 1006}, 075 (2010)
ÊÊ[arXiv:1002.3805 [hep-th]].
ÊÊ

\item{[37]}
ÊÊM.~B.~Green, S.~D.~Miller, J.~G.~Russo and P.~Vanhove,
ÊÊ{\it Eisenstein series for higher-rank groups and string theory
amplitudes},
ÊÊarXiv:1004.0163 [hep-th].
ÊÊ

\item{[38]}
ÊÊB.~Pioline,
ÊÊ{\it R**4 couplings and automorphic unipotent representations},
ÊÊJHEP {\bf 1003}, 116 (2010)
ÊÊ[arXiv:1001.3647 [hep-th]].
ÊÊ

\item{[39]}
ÊÊF.~Gubay, N.~Lambert, P.~West,
ÊÊ{\it Constraints on Automorphic Forms of Higher Derivative Terms
from Compactification},
ÊÊJHEP {\bf 1008}, 028 (2010).
ÊÊ[arXiv:1002.1068 [hep-th]]

\item{[40]}
ÊÊN.~Berkovits,
ÊÊ{\it New higher-derivative R**4 theorems},
ÊÊPhys.\ Rev.\ Lett. Ê{\bf 98}, 211601 (2007)
ÊÊ[arXiv:hep-th/0609006].
ÊÊ

\item{[41]}
ÊÊP.~C.~West,
ÊÊ{\it Hidden superconformal symmetry in M theory},
ÊÊJHEP {\bf 0008}, 007 (2000)
ÊÊ[arXiv:hep-th/0005270].
ÊÊ

ÊÊ
\item{[42]}
ÊÊI.~Schnakenburg and P.~C.~West,
ÊÊ{\it Kac-Moody symmetries of 2B supergravity},
ÊÊPhys.\ Lett. ÊB {\bf 517}, 421 (2001)
ÊÊ[arXiv:hep-th/0107181].
ÊÊ

ÊÊÊÊÊÊ
\item{[43]}
ÊÊP.~C.~West,
ÊÊ{\it E(11) and M theory},
ÊÊClass.\ Quant.\ Grav. Ê{\bf 18}, 4443 (2001)
ÊÊ[arXiv:hep-th/0104081].
ÊÊ

\item{[44]}
ÊÊF.~Riccioni and P.~C.~West,
ÊÊ{\it The E(11) origin of all maximal supergravities},
ÊÊJHEP {\bf 0707}, 063 (2007)
ÊÊ[arXiv:0705.0752 [hep-th]].

\item{[45]}
ÊÊF.~Riccioni and P.~C.~West,
ÊÊ{\it E(11)-extended spacetime and gauged supergravities},
ÊÊJHEP {\bf 0802}, 039 (2008)
ÊÊ[arXiv:0712.1795 [hep-th]].

\item{[46]}
ÊÊF.~Riccioni and P.~West,
ÊÊ{\it Local E(11)},
ÊÊJHEP {\bf 0904}, 051 (2009)
ÊÊ[arXiv:0902.4678 [hep-th]].
ÊÊ

\item{[47]}
ÊÊF.~Riccioni, D.~Steele and P.~West,
ÊÊ{\it The E(11) origin of all maximal supergravities: The Hierarchy of
ÊÊfield-strengths},
ÊÊJHEP {\bf 0909}, 095 (2009)
ÊÊ[arXiv:0906.1177 [hep-th]].
ÊÊ

\item{[48]}
ÊÊA.~Kleinschmidt and P.~C.~West,
ÊÊ{\it Representations of G+++ and the role of space-time},
ÊÊJHEP {\bf 0402}, 033 (2004)
ÊÊ[arXiv:hep-th/0312247].
ÊÊ

\item{[49]}
ÊÊP.~C.~West,
ÊÊ{\it E(11) origin of brane charges and U-duality multiplets},
ÊÊJHEP {\bf 0408}, 052 (2004)
ÊÊ[arXiv:hep-th/0406150].
ÊÊ

\item{[50]}
ÊÊP.~C.~West,
ÊÊ{\it Brane dynamics, central charges and E(11)},
ÊÊJHEP {\bf 0503}, 077 (2005)
ÊÊ[arXiv:hep-th/0412336].
ÊÊ

\item{[51]}
ÊÊP.~P.~Cook and P.~C.~West,
ÊÊ{\it Charge multiplets and masses for E(11)},
ÊÊJHEP {\bf 0811}, 091 (2008)
ÊÊ[arXiv:0805.4451 [hep-th]].
ÊÊ

\item{[52]}
ÊÊN.~A.~Obers, B.~Pioline, E.~Rabinovici,
ÊÊ{\it M theory and U duality on T**d with gauge backgrounds},
ÊÊNucl.\ Phys. Ê{\bf B525}, 163-181 (1998).
ÊÊ[hep-th/9712084]
%

\item{[53]}
ÊÊP.~West,
ÊÊ{\it Generalised space-time and duality},
ÊÊPhys.\ Lett. ÊB {\bf 693}, 373 (2010)
ÊÊ[arXiv:1006.0893 [hep-th]].
ÊÊ
%

%
\item{[54]}
ÊÊM.~R.~Gaberdiel, D.~I.~Olive and P.~C.~West,
ÊÊ{\it A Class of Lorentzian Kac-Moody algebras},
ÊÊNucl.\ Phys. ÊB {\bf 645}, 403 (2002)
ÊÊ[arXiv:hep-th/0205068].
ÊÊ

\item{[55]}
ÊÊP.~C.~West,
ÊÊ{\it The IIA, IIB and eleven-dimensional theories and their common
E(11) origin},
ÊÊNucl.\ Phys. Ê{\bf B693}, 76-102 (2004).
ÊÊ[hep-th/0402140].

\item{[56]}
ÊÊP.~C.~West,
ÊÊ{\it E(11), SL(32) and central charges},
ÊÊPhys.\ Lett. Ê{\bf B575}, 333-342 (2003).
ÊÊ[hep-th/0307098].
\end